\documentclass[aps,prc,showpacs,showkeys,nofootinbib]{revtex4}

\usepackage{graphicx}
\usepackage{dcolumn}
\usepackage{amsthm}
\usepackage{bm}

\usepackage{isotope}
\usepackage{feyn}

\newcommand{\definition}{\textit}

\newcommand{\sigmabf}{\mbox{\boldmath $\sigma$}}

\newcommand{\rhobf}{\mbox{\boldmath $\rho$}}
\newcommand{\rhobfsm}{\small \mbox{\boldmath $\rho$}}

\begin{document}

\title{Manifestation of important role of nuclear forces in emission of photons in scattering of pions off nuclei
}


\author{Sergei~P.~Maydanyuk$^{(1,2)}$}%
\email{maidan@kinr.kiev.ua}%
\author{Peng-Ming~Zhang$^{1}$}%
\email{zhpm@impcas.ac.cn} %
\author{Li-Ping~Zou$^{1}$}%
\email{zoulp@impcas.ac.cn} %
\affiliation{$(1)$Institute of Modern Physics, Chinese Academy of Sciences, Lanzhou, 730000, China}
\affiliation{$(2)$Institute for Nuclear Research, National Academy of Sciences of Ukraine, Kiev, 03680, Ukraine}

\date{\small\today}


\begin{abstract}
Bremsstrahlung of photons emitted during the scattering of $\pi^{+}$-mesons off nuclei is studied for the first time.
Role of interactions between $\pi^{+}$-mesons and nuclei in the formation of the bremsstrahlung emission is analyzed in details.
We discover essential contribution of emitted photons from nuclear part of Johnson-Satchler potential to the full spectrum, in contrast to the optical Woods-Saxon potential.
%
%
We observe unusual essential influence of the nuclear part of both potentials on the spectrum at high photon energies.
This phenomenon 
opens a new experimental way to study and check non-Coulomb and nuclear interactions between pions and nuclei via measurements of the emitted photons.
We provide predictions of the bremsstrahlung spectra for pion scattering off $^{44}{\rm Ca}$.
\end{abstract}


\maketitle

\section{Introduction
\label{sec.introduction}}

The nature of pion-nucleus interaction in the energy region from 0 to 1~GeV has been the subject of considerable theoretical and experimental investigations~\cite{Kluge.1991.RepPP}.
Motivation of such researches is using pions in nuclear reactions as a probe to obtain deeper understanding of nuclear structure
\cite{Hirata.1979.AP,Oset.1982.PRep,Ericson_Weise.1988.book}.
%
%
Particular attention has been paid to understanding the discrete and (low-lying) collective states of nuclei in elastic, inelastic and quasielastic scattering, resonant scattering (with and without polarized nuclei), single and double charge exchange reactions, knock-out reactions and absorption reactions.
Moreover, a lot of interests were also focused on understanding the pion-nucleon $\Delta_{3,3}$ ($J=\frac{3}{2}$, $T=\frac{3}{2}$) resonance and its role in the pion-nucleus interactions.
%
The experimental measurements have been performed
at CERN Synchro-cyclotron~\cite{Binon.1970.NPB},
in $M 13$ pion channel using the quadrupole-quadrupole-dipole pion spectrometer at TRIUMF (Vancouver)~\cite{Sobie.1984.PRC}
(for example, see a detailed description of the channel and the spectrometer in Refs.~\cite{Oram.1981.NIM,Sobie.1984.NIM}),
at 7 GeV proton synchrotron NIMROD at Rutherford Laboratory~\cite{Clough.1974.NPB},
in pion channel and spectrometer (EPICS) at Clinton P. Anderson Meson Physics Facility (LAMPF, Los Alamos)~\cite{Clough.1974.NPB,Morris.1981.PRC,Boyer.1981.PRC,Preedom.1981.PRC},
at pion spectrometer facility at Swiss Institute for Nuclear Research (SIN, Switzerland)~\cite{Preedom.1979.NPA,Albanese.1980.NPA}.
Experimental program (for wide range of experiments) has been proposed at SIN 
to study the pion-nucleus interactions in systematic way (see Ref.~\cite{Albanese.1979.NIM} for details).
%
%
Until now, we have experimental data of cross-sections for
$^{6,7}{\rm Li}$, $^{9}{\rm Be}$, $^{12}{\rm C}$, $^{16}{\rm O}$, $^{28}{\rm Si}$, $^{32,34}{\rm S}$, $^{40, 42,44,48}{\rm Ca}$, $^{54}{\rm Fe}$, $^{90}{\rm Zr}$, $^{208}{\rm Pb}$, etc.
at the incident pion energy $T_{\pi}$ from 30~MeV up to 860~MeV~\cite{
Clough.1974.NPB,
Preedom.1981.PRC,
Sobie.1984.PRC,
Morris.1981.PRC,
Binon.1970.NPB,
Albanese.1980.NPA,
Preedom.1979.NPA,
Boyer.1981.PRC,
Boyer.1984.PRC}.


Optical model \cite{Preedom.1979.NPA,Preedom.1981.PRC,Sobie.1984.PRC,Khallaf.2000.PRC,Akhter.2001.JPG,Khallaf.2002.PRC},
$\alpha$-cluster model~\cite{Ebrahim.2011.BrJP},
DWIA collective model~\cite{Preedom.1979.NPA,Morris.1981.PRC,Boyer.1981.PRC,Boyer.1984.PRC},
microscopical form-factor model~\cite{Boyer.1981.PRC}
have been used to analyze (and predict in some cases) total and differential cross section measurements.
The parameters in pion-nucleus interaction in these models were determined by fitting procedures.
The ground-state neutron density root-mean-square radii (for $^{40,42,44,48}{\rm Ca}$ and $^{54}{\rm Fe}$) was estimated~\cite{Boyer.1984.PRC}.
Neutron and proton multipole matrix elements for excited states (of $^{42,44,48}{\rm Ca}$, see Ref.~\cite{Boyer.1981.PRC}) have also been experimentally tested.
Deformation lengths in inelastic scattering were extracted from experimental data
[lengths for ($2^{+}$; 4.44~MeV) and ($3^{-}$; 9.64~MeV) states of $^{12}{\rm C}$~\cite{Ebrahim.2011.BrJP}, 
etc.].
These characteristics confirm perspectives of the study on pion-nucleus interaction (reactions), when electromagnetic measurements do not exist.

Theoretically, in describing the pion-nucleus interaction, there are two prevailing approaches.
The first one was introduced by Satchler in Ref.~\cite{Satchler.1992.NPA} and developed in frameworks of optical model with non-relativistic Schr\"{o}dinger equation and phenomenological local potential of Woods-Saxon shape
(see Ref.~\cite{Akhter.2001.JPG}).
The second one, which was introduced by Johnson and Satchler in Ref.~\cite{Johnson.1996.AP} on the basis Klein-Gordon equation, uses a nonlocal Kisslinger-type potential at low and resonance energies.
Via Krell-Ericson transformation, the later formalism is modified to the optical model formalism with local potentials and transformed wave functions (see Ref.~\cite{Khallaf.2000.PRC,Khallaf.2002.PRC}).
Note that key point in effective description of pion-nucleus interactions is proper nuclear interaction which is deeply investigated.


Bremsstrahlung emission of photons accompanying nuclear reactions has been attracting a lot of interests for a long time
(see reviews~\cite{Pluiko.1987.PEPAN,Kamanin.1989.PEPAN} and books~\cite{Amusia_Buimistrov.1987,Amusia.1990}).
This is because photons provide independent information about the nuclear processes,
and it can be used as an independent probe for nuclear interactions.
%
%
However, the nature of emission of hard photons in the nucleus-nucleus collisions is open question, because of the complicated many-nucleon interactions.
Because of this, before attempts to resolve the fully many-nucleon problem of nuclear scattering with emission of photons, people go to a simpler bremsstrahlung calculations with proton chosen as one scattering nuclear object.
However, even for the proton-nucleus scattering, a clear understanding of influence of nuclear interactions on the spectrum of photons has not been obtained~\cite{Kopytin.1997.YF}.

In this paper, we find that our calculated bremsstrahlung spectra are sufficiently sensitive to the shape of this type of the potential.
Thus, we propose that the emitted bremsstrahlung photons can be a suitable tool to study the non-Coulomb interactions (i.e., nuclear forces, Coulomb corrections, etc.)
in the scattering of the positively charged pions off nuclei.
However, the bremsstrahlung emission by real pions scattered off nuclei has never been studied, both theoretically and experimentally.
Note that emission of the coherent photons by virtual pions, which are produced in nuclear matter during the proton - nucleus scattering,
was studied~\cite{Gil.1998.PLB} in the $\Delta$-resonance energy region.
However, in the formation of photons during this reaction the Coulomb forces outside the nucleus-target (which give the largest contribution to the coherent and incoherent
bremsstrahlung emission in the pion-nucleus scattering) are not included in
consideration%
\footnote{In order to obtain convergence in calculations of the matrix elements of the bremsstrahlung emission during scattering of the real pion off nucleus,
we must include space region of integration with external boundary up to atomic shells of the nucleus.
From here, one can see the essential role of the Coulomb forces in estimating bremsstrahlung spectral.
In Ref.~\cite{Gil.1998.PLB} essentially different process of emission of photons was investigated, without such a Coulomb contribution.}.
In this paper we report our analysis on the question above.
We find that nuclear interactions play an important role in the formation of bremsstrahlung spectra,
which can be tested experimentally (at high energy region of photons) via further measurements of bremsstrahlung photons.

\section{Model
\label{sec.2}}

In construction of our bremsstrahlung formalism
(see for proton-nucleus scattering and proton decay~\cite{Maydanyuk.2011.JPG,Maydanyuk.2012.PRC,Maydanyuk_Zhang.2015.PRC},
$\alpha$ decay~\cite{Maydanyuk.2003.PTP,Maydanyuk.2006.EPJA,Maydanyuk.2008.EPJA,Giardina.2008.MPLA,
Maydanyuk.2009.NPA,Maydanyuk.2009.TONPPJ,Maydanyuk.2009.JPS,Maydanyuk_Zhang_Zou.2016.PRC},
spontaneous fission~\cite{Maydanyuk.2010.PRC},
ternary fission~\cite{Maydanyuk.2011.JPCS})
we use the following logic.
We begin formalism with Dirac equation for one nucleon.
Then, we generalize this equation for description of many nucleons
(we take radius-vector of center of mass of the full system as a summation of all radius-vectors of nucleons).
As the next step, we apply approximation (known in QED for one fermion), which allows to obtain generalization of Pauli equation for many nucleons, as first approximation [we obtain a logic how to determine the next approximations (p.~33--35 in Ref.~\cite{Ahiezer.1981}), what is reduced mass, relation of such a formalism with relativistic classical mechanics with action (p.~35--36 in Ref.~\cite{Ahiezer.1981}), etc.].
In this step, we introduce vector potentials of the electromagnetic field (in the standard way of QED via gauge) which describe emission of photons by each nucleon.
This formalism should be transformed to the known Pauli equation after imposing the needed approximations.
We formulate initially the hamiltonian of the full many-nucleon nuclear system in laboratory frame.
We use space variables for each nucleon.
On its basis, the full operator of emission of the evolving nuclear system is defined in laboratory frame.
Simply, relative acceleration between nucleons forms the emission of bremsstrahlung photons.
Thus, we need in relative distances between nucleons for analysis, and we rewrite the full hamiltonian and full wave function via these relative distances.
As result, we obtain term in operator of emission corresponding to motion of full nuclear system in laboratory frame (for example, see first term in Eq.~(8) in Ref.~\cite{Maydanyuk_Zhang.2015.PRC} for proton-nucleus scattering).
In construction of the wave function of the full nuclear system, we separate factor describing motion of full nuclear system (for example, see (10) in Ref.~\cite{Maydanyuk_Zhang.2015.PRC} for the proton-nucleus scattering).
And we separate factor of relative motion, as we estimate that relative motion between two nuclear fragments gives the largest contribution to the full bremsstrahlung spectrum (we need in this factor in calculations).

In this paper we generalize this formalism for the pion-nucleus scattering.
For the first estimations of the bremsstrahlung photons which are emitted during a scattering of pions off nuclei, we put main forces to determinate a leading contribution of emission to the full spectrum. Such a term is based on relative motion of the nucleus and pion.
We shall not study internal structure of these two object in this paper,
as they should give smaller effects (but will be estimated as the nest step in this research).%
\footnote{We explained the observed hump-shaped plateau in the experimental bremsstrahlung spectra in the proton-nucleus scattering in the intermediate and high-energy regions, by not-minor incoherent emission of photons (formed by interactions between spinor properties of individual nucleons and their momenta) \cite{Maydanyuk_Zhang.2015.PRC}.
Such a behavior is described as an addition to the leading contribution of the emitted coherent photons. 
Thus, in this paper, we neglect such a non-coherent effect (and, so, spinor properties of nucleons).
}



Emission of the bremsstrahlung photons can be introduced to the formalism of the $\pi^{\pm}$-nucleus scattering via Coulomb gauge for each electromagnetic charge in the system
as
%
$\mathbf{p}_{i} \to \mathbf{p}_{i} - \displaystyle\frac{z_{i}\, e}{c}\, \mathbf{A} (\mathbf{r}_{i}, t)$,
%
where
$\mathbf{p}_{i} = - i\hbar\, \nabla_{i}$ is the momentum of pion or nucleon with number $i$,
$\mathbf{A} (\mathbf{r}_{i}, t)$ is the vector potential of the electromagnetic field formed by motion of pion or nucleon with number $i$,
$z_{i}$ is the electromagnetic charge of pion or nucleon with number $i$.
The modified Hamiltonian is written as $\hat{H} = \hat{H}_{0} + \hat{H}_{\gamma}$,
where $\hat{H}_{\gamma}$ is a new operator describing emission of photons.
A leading part of emission operator of the system composed of $\pi^{\pm}$ and nucleus in the laboratory frame 
(neglecting terms at $\mathbf{A}_{i}^{2}$, $A_{i,0}$ and spinor terms,
see Appendix~\ref{sec.app.1}
for detailed derivation of operator of emission of evolving system of nucleons and pion, and needed definitions) has a form:
%
%
\begin{equation}
\begin{array}{lcl}
  \hat{H}_{\gamma} =
  -\,e\, \sqrt{\displaystyle\frac{2\pi\hbar}{w_{\rm ph}}}\,
    \displaystyle\sum\limits_{\alpha=1,2} \mathbf{e}^{(\alpha),*}\;
    e^{-i \mathbf{k}\, \bigl[\mathbf{R} - \displaystyle\frac{m_{\pi}}{m_{A}+m_{\pi}}\:
    \mathbf{r}\bigr]}\;
  \Biggl\{
    \displaystyle\frac{1}{m_{A} + m_{\pi}}\;
    \biggl[
      e^{-i \mathbf{k}\, \mathbf{r}}\; z_{\pi} +
      \displaystyle\sum\limits_{j=1}^{A} z_{Aj}\; e^{-i \mathbf{k}\, \rhobfsm_{Aj}}
    \biggr]\; \mathbf{P}\; +\; \\
  + \;
    \biggl[
      e^{-i \mathbf{k}\, \mathbf{r}}\; \displaystyle\frac{z_{\pi}}{m_{\pi}}\; -
      \displaystyle\frac{1}{m_{A}}
      \displaystyle\sum\limits_{j=1}^{A}
        z_{Aj}\; e^{-i \mathbf{k}\, \rhobfsm_{Aj}}
    \biggr]\; \mathbf{p}\; + \;
  \displaystyle\sum\limits_{j=1}^{A-1}
      \displaystyle\frac{z_{Aj}}{m_{Aj}}\; e^{-i \mathbf{k}\,
      \rhobfsm_{Aj}}\, \mathbf{\tilde{p}}_{Aj}\; -\;
  \displaystyle\frac{1}{m_{A}}
  \biggl[
    \displaystyle\sum\limits_{j=1}^{A}
      z_{Aj}\; e^{-i \mathbf{k}\, \rhobfsm_{Aj}}
  \biggr]\; \displaystyle\sum_{k=1}^{A-1} \mathbf{\tilde{p}}_{Ak}
  \Biggr\},
\end{array}
\label{eq.2.3.2}
\end{equation}
where
$\mathbf{P} = -i\hbar\, \mathbf{d}/\mathbf{dR}$,
$\mathbf{p} = -i\hbar\, \mathbf{d}/\mathbf{dr}$,
$\mathbf{\tilde{p}}_{Aj} = -i\,\hbar\, \mathbf{d}/\mathbf{d}\rhobf_{Aj}$,
$\mathbf{R}$ are coordinates of center of mass of complete nuclear system of the nucleus and pion,
$\mathbf{r}$ is relative distance between center-of-mass of the nucleus and pion,
$\rhobf_{Aj}$ is relative distance between center-of-mass of the nucleus and nucleon with number $j$ of this nucleus,
$\mathbf{R}$, $\mathbf{r}$, $\rhobf_{Aj}$ are defined in (\ref{eq.app.2.1.3}),
star denotes complex conjugation,
$m_{i}$ and $z_{i}$ are mass and electromagnetic charge of nucleon with number $i$ ($i$=1 \ldots A)
or mass and electromagnetic charge of $\pi^{\pm}$-meson ($i=A+1$),
$m_{\pi}$ and $m_{A}$ are masses of $\pi^{\pm}$ and nucleus,
$\mathbf{e}^{(\alpha)}$ are unit vectors of polarization of the photon emitted,
$\mathbf{k}$ is the wave vector of the photon,
$w_{\rm ph} = k\, c = \bigl| \mathbf{k}\bigr| c$.
%
%
%
The diagrammatic representation of the emission of the bremsstrahlung photons in the pion-nucleus scattering is depicted in Fig.~\ref{fig.1.add}.
\begin{figure}[htbp]
\centering
\begin{equation}
\begin{array}{lll}
  \Diagram{\vertexlabel^{\pi^{\prime}} \\ hd \\ & g\vertexlabel_{k} \\ \vertexlabel_{\pi} hu\\} 
  \; + \; \displaystyle\sum\limits_{i=1}^{Z=20} \Diagram{\vertexlabel^{p_{i}^{\prime}} \\ fd \\ & g\vertexlabel_{k} \\ \vertexlabel_{p_{i}} fu\\}
  \; + \; \displaystyle\sum\limits_{j=1}^{A-Z=24} \Diagram{\vertexlabel^{n_{j}^{\prime}} \\ fd \\ & g\vertexlabel_{k} \\ \vertexlabel_{n_{j}} fu\\}
\end{array}
\end{equation}
\vspace{0mm}
\caption{\small 
Diagrammatic representation of the emission of the bremsstrahlung photons during the scattering of the $\pi^{+}$-meson off the \isotope[44]{Ca} nucleus.
Dashed and solid lines refer to pions and nucleons, respectively.
Wavy lines indicate the outgoing bremsstrahlung photons.
\label{fig.1.add}}
\end{figure}


The Hamiltonian of the evolving system of the nucleus and $\pi^{\pm}$-meson
is
$\hat{H}_{0} = -\, \displaystyle\frac{\hbar^{2}}{2m}\, \triangle + V_{N}(\mathbf{r}) + V_{C}(\mathbf{r})$,
%
%
where
$V_{C}$ is Coulomb potential of interactions between the pion and nucleus,
$V_{N}$ is potential of non-Coulomb interactions between the pion and nucleus,
$m$ is reduced mass of the pion and nucleus~\cite{Khallaf.2000.PRC}.
The component $V_{N}$ can be written
in the local Kisslinger-type form~\cite{Khallaf.2000.PRC,Khallaf.2002.PRC}
along the formalism of Johnson and Satchler~\cite{Johnson.1996.AP}
as (we add upper indexes \emph{``JS''} and \emph{``WS''} to denote type of formalism)
\begin{equation}
\begin{array}{lcl}
  V_{N}^{(JS)}(\mathbf{r}) & = & U_{N}(r) + \Delta\, U_{C}(r), \\
%
  U_{N}(r) & = &
  \displaystyle\frac{(\hbar c)^{2}} {2w}\;
  \biggl\{
    \displaystyle\frac{q(r)} {1 - \alpha(r)} -
    \displaystyle\frac{k^{2}\alpha(r)} {1 - \alpha(r)} -
    \displaystyle\frac{\nabla^{2}\alpha(r)} {2\,(1 - \alpha(r))} -
    \Bigl[ \displaystyle\frac{\nabla \alpha(r)} {2\,(1 - \alpha(r))} \Bigr]^{2}
  \biggr\}, \\
%
  \Delta U_{C}(r) & = &
  \displaystyle\frac{\alpha(r) V_{C}(r) - V_{C}^{2}(r) / 2w} {1 - \alpha(r)},
\end{array}
\label{eq.2.2.4}
\end{equation}
or in the local Woods-Saxon form along the optical model formalism~\cite{Akhter.2001.JPG} as%
\footnote{In this paper we restrict ourselves by only real part of potential in calculations of the bremsstrahlung spectra.
Motivations of such calculations are the following.
The imaginary part of potential is related with internal (inelastic) mechanisms inside the nucleus, which are essentially more difficult for analysis.
Inclusion of the inelastic mechanisms will give additional new questions in this tasks (for example, a role of such mechanisms in the scattering in dependence on energy of emitted photons;
such mechanisms can be related with producing of incoherent emission, etc.).
Thus, we want initially to construct proper basis for description of the bremsstrahlung in the elastic scattering.
In the next step, it is sensible to estimate inelastic mechanisms, using already constructed basis for bremsstrahlung in the elastic scattering.
We find the not small role of the nuclear part of real potential in determination of the spectra at high energies of photons is an interesting new effect.
Importance of studying this effect will be after adding imaginary part of the potential to calculations.
}
\begin{equation}
\begin{array}{llll}
  V_{N}^{(WS)}(\mathbf{r}) = - U\, f(r) - i\, W\, g(r), &
  f(r) = [1 + \exp((r - R_{u})/a_{u})]^{-1}, &
  g(r) = [1 + \exp((r - R_{w})/a_{w})]^{-1}.
\end{array}
\label{eq.2.2.5}
\end{equation}
Here,
$w$ is total energy of pion in center-of-mass frame,
$q(r)$ and $\alpha(r)$ are results of $s$-wave part and $p$-wave part of pion--nucleon interactions (see Ref.~\cite{Khallaf.2000.PRC,Johnson.1996.AP} for details, reference).
$U$ and $W$ are strengths of the potential (\ref{eq.2.2.5}), 
$R_{u}$ and $R_{w}$ are radius-parameters of the potential (\ref{eq.2.2.5}),
$a_{u}$ and $a_{w}$ are diffuseness (see Table~2 in Ref.~\cite{Akhter.2001.JPG}).


The leading matrix element of emission is
(see Appendix~\ref{sec.app.2}
for its derivation and definition of the wave function of pion-nucleons system)
%
%
\begin{equation}
\begin{array}{ll}
  \langle \Psi_{f} |\, \hat{H}_{\gamma} |\, \Psi_{i} \rangle_{1} =
  -\,\displaystyle\frac{e}{m}\:
  \sqrt{\displaystyle\frac{2\pi\hbar}{w_{\rm ph}}}\;
    \displaystyle\sum\limits_{\alpha=1,2} \mathbf{e}^{(\alpha),*}\;
    \biggl\langle\: \Phi_{\rm \pi-nucl, f} (\mathbf{r})\: \biggl|\,
      Z_{\rm eff}(\mathbf{k}, \mathbf{r})\: e^{-i\,\mathbf{kr}}\: \mathbf{p}\;
    \biggr|\: \Phi_{\rm \pi-nucl, i} (\mathbf{r})\: \biggr\rangle,
\end{array}
\label{eq.2.5.8}
\end{equation}
where we introduce the \emph{effective charge of the $\pi^{\pm}$--nucleus system} as
\begin{equation}
\begin{array}{lcl}
  Z_{\rm eff}(\mathbf{k}, \mathbf{r}) & = &
    e^{i\,\mathbf{kr}\, \displaystyle\frac{m_{\pi}}{m_{A}+m_{\pi}}}\:
    \biggl\{
      \displaystyle\frac{m_{A}\, z_{\pi}}{m_{A}+m_{\pi}} -
      e^{i\,\mathbf{kr}}\:
      \displaystyle\frac{m_{\pi}\, Z_{\rm A}(\mathbf{k})}{m_{A}+m_{\pi}}
    \biggr\}
\end{array}
\label{eq.2.5.3}
\end{equation}
the \definition{charged form-factor of the nucleus} as
\begin{equation}
  Z_{\rm A} (\mathbf{k}) =
  \Bigl\langle \psi_{\rm nucl, f} (\rhobf_{A1} \ldots \rhobf_{AA})\: \Bigl|\;
    \displaystyle\sum\limits_{j=1}^{A}
      z_{Aj}\:
      e^{-i \mathbf{k} \rhobfsm_{Aj}}\:
  \Bigr|\: \psi_{\rm nucl, i} (\rhobf_{A1} \ldots \rhobf_{AA})\, \Bigr\rangle.
\label{eq.2.5.4}
\end{equation}
%
%
%
%
Here,
$\Psi_{s}$ is the wave function of the full pion-nuclear system,
$s = i$ or $f$ (indexes $i$ and $f$ denote the state before emission of photon and the state after emission of photon),
$\Phi_{\rm \pi-nucl,s} (\mathbf{r})$ is function describing relative motion (with tunneling) of the $\pi^{\pm}$ related to the nucleus (without description of internal relative motions of nucleons inside the nucleus),
$\psi_{\rm nucl,s}(\beta)$ is the many-nucleon function describing internal states of nucleons in the nucleus
(it determines space states on the basis of relative distances $\rhobf_{1}$ \ldots $\rhobf_{A}$ of nucleons of the nucleus concerned with its center-of-masses, and spin-isospin states also),
$\beta_{A}$ is a set of numbers $1 \ldots A$ of nucleons of the nucleus.

The simplest determination of the matrix element is obtained, if
(1) to apply dipole approximation for the effective charge of the full system, and
(2) to neglect relative displacements of nucleons inside the nucleus in calculations of the form-factor
(see Appendix~\ref{sec.app.3}
for derivation of the matrix elements).
At such an approximation, 
we define the cross section of the emitted photons in the laboratory frame
in frameworks of formalism given in Ref.~\cite{Maydanyuk.2012.PRC} and we do not repeat it in this paper [see Eq.~(49) in that paper]:
%
%
%
%
\begin{equation}
\begin{array}{ccl}
  \displaystyle\frac{d^{2}\,\sigma (\theta_{f})}{dw_{\rm ph}\,
  d\cos{\theta_{f}}} & = &
    \displaystyle\frac{e^{2}}{2\pi\,c^{5}}\:
      \displaystyle\frac{w_{\rm ph}\,E_{i}}{m^{2}\,k_{i}} \;
      \biggl\{p_{\rm el}\, \displaystyle\frac{d\, p_{\rm el}^{*}(\theta_{f})}{d\,\cos{\theta_{f}}}
      + {\rm c. c.} \biggr\},
\end{array}
\label{eq.2.6.1}
\end{equation}
where c.~c. is complex conjugation
%
(see Appendix~\ref{sec.app.3}
for derivation of the matrix elements).
%
%
We calculate radial wave functions numerically concerned with the chosen potential
of interaction between the $\pi^{\pm}$ and the spherically symmetric core
[see Eqs.~(\ref{eq.2.2.4}) for the nuclear potential and corrections along the the Johnson and Satchler formalism,
see Eqs.~(\ref{eq.2.2.5}) for the nuclear potential along the Woods-Saxon formalism].
\section{Discussions
\label{sec.results}}

\begin{figure}[htbp]
\centerline{\includegraphics[width=85mm]{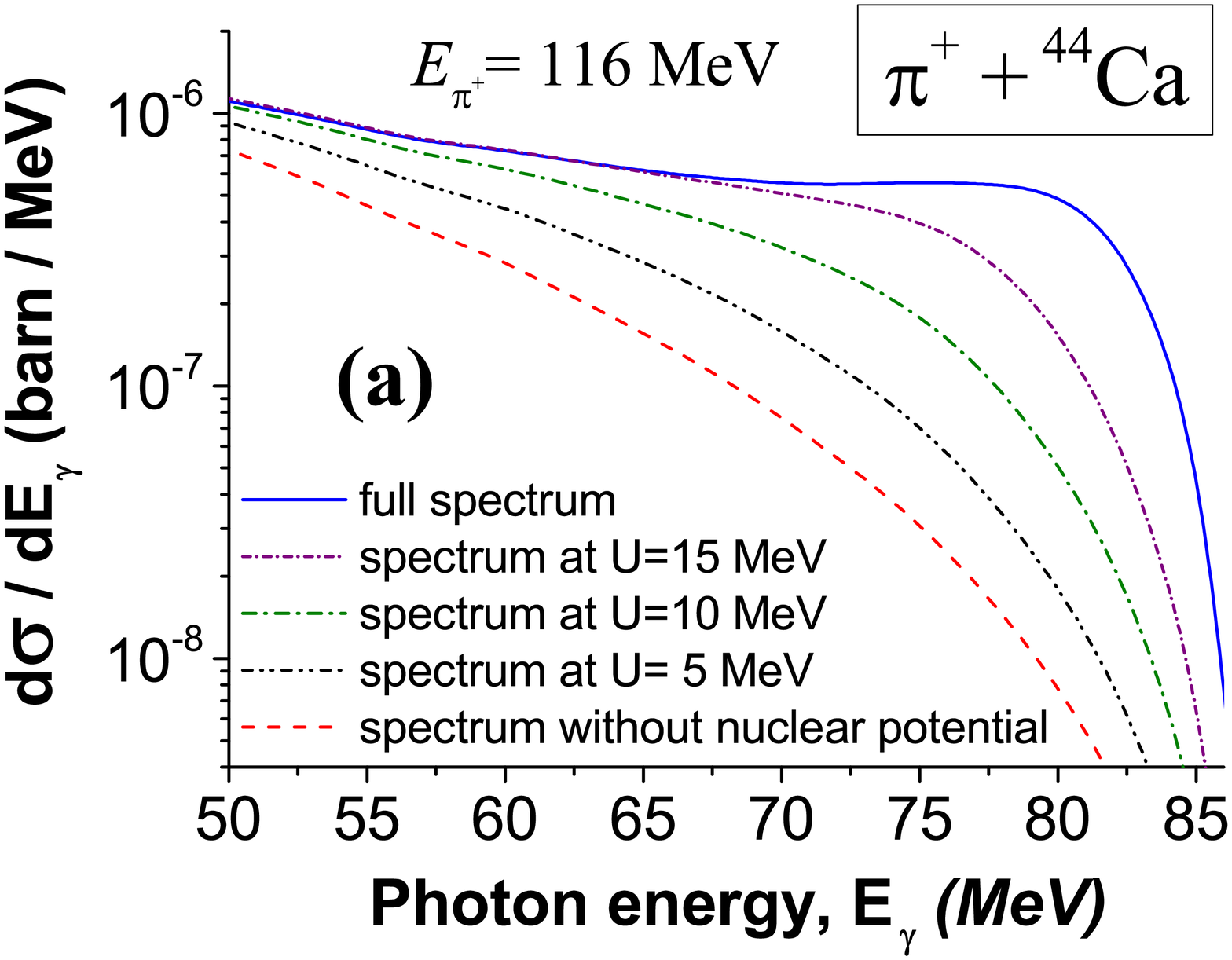}
\hspace{-1mm}\includegraphics[width=85mm]{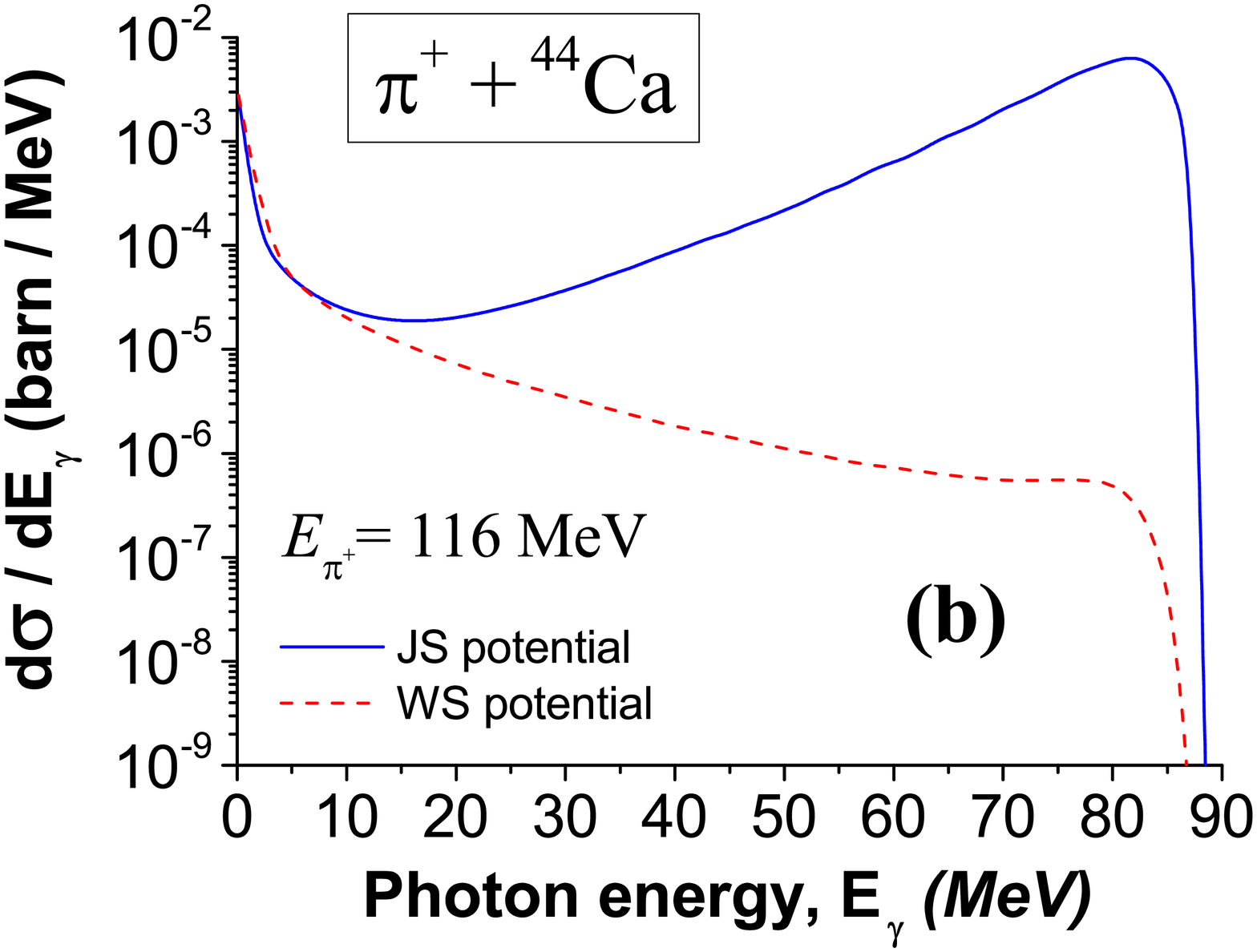}}
\vspace{-6mm}
\caption{\small (Color online)
The bremsstrahlung cross sections of photons emitted during the scattering of the $\pi^{+}$-mesons off
the $^{44}{\rm Ca}$ nucleus at the bombarding energy of 116~MeV
[parameters of calculations:
the potential $V_{N}^{(WS)}$ and $V_{N}^{(JS)}$ are defined in Eqs.~(\ref{eq.2.2.5}) and Eqs.~(\ref{eq.2.2.4}),
parameters of these potentials are taken in Ref.~\cite{Akhter.2001.JPG} and Table~1 in Ref.~\cite{Johnson.1996.AP},
the angle $\theta_{\gamma}$ between directions of the photon emission and the $\pi$-mesons motion is $90^{\circ}$%
].
%
[Panel a]:
The bremsstrahlung spectra, where interactions between $\pi^{\pm}$-mesons and nuclei are described by the Woods-Saxon optical model.
We show changes of the spectrum in dependence on strength $U$ of the nuclear potential~(\ref{eq.2.2.5}).
One can see stability of calculations.
%
[Panel b]:
The bremsstrahlung spectra, where interactions between $\pi^{\pm}$-mesons and nuclei are defined in the Johnson-Satchler formalism (see blue solid line)
and the Woods-Saxon optical model formalism (see red dashed line).
One can see principally different behavior between the spectra.
This difference is because two theories used in the basis of calculations of these bremsstrahlung spectra are essentially different
(the Coulomb interactions and Centrifugal potential terms are the same):
the Johnson-Satchler formalism is based on the relativistic Klein-Gordon equation, while the optical model formalism is based on non-relativistic Shr\"{o}dinger equation.
Note that both spectra obtained by the Johnson-Satchler formalism and the optical model formalism with Woods-Saxon potential are coincide at low photon energies that confirms our formalism and calculations (calculations are not normalized on any data point).
\label{fig.1}}
\end{figure}

We apply the approach presented above to estimate the spectra of the bremsstrahlung photons emitted during the scattering of the $\pi^{+}$-meson off nuclei.
We begin with calculations of the bremsstrahlung photons where interactions between $\pi^{+}$-mesons and nuclei are defined in the optical model formalism with the Woods-Saxon nuclear potential.
We estimate influence of the nuclear part of potential on the spectrum.
Such calculations of the bremsstrahlung cross-sections in scattering of $\pi^{+}$-mesons of the $^{44}{\rm Ca}$ nuclei
at the bombarding meson energy $E_{\pi^{+}}=116$~MeV are presented in Fig.~\ref{fig.1}~(a).
One can see clear difference between the spectra at high energy region of the emitted photons caused by variations of strength $U$ of nuclear potential~(\ref{eq.2.2.5}).
This is manifestation of important role of nuclear interactions in the formation of emission of photons.
This phenomenon is a more visible in the high energy region.
We analyze how much the bremsstrahlung emission is changed, if the interaction between pions and nuclei are defined in the Johnson-Satchler formalism.
Such calculations for the $^{44}{\rm Ca}$ nucleus at the bombarding energy $E_{\pi^{\pm}}=116$ MeV are presented in Fig.~\ref{fig.1}~(b)
in comparison with previous results given in Fig.~\ref{fig.1}~(a).
One can see a principally different behavior between the spectra.

At present, there is no experimental information about bremsstrahlung photons emitted during the scattering of pions off nuclei
(we have even not found any idea about these investigations in the literature).
Thus, there is no experimental basis for conclusion about the realistic nuclear parameters of the pion-nucleus potential using analysis of the bremsstrahlung emission.
At the same time, such photons should be emitted in this reaction.
We show sensitivity of the bremsstrahlung spectra to parameter $U$.
We conclude that analysis of the bremsstrahlung emission has good perspective to obtain estimations of this parameter from experimental study.

Without analysis of bremsstrahlung photons, some investigations of the interacting potential in frameworks of optical model were given in Ref.~\cite{Akhter.2001.JPG}. We use results of work of those people.
Here, there are the following parameters for reaction $\pi^{+} + ^{44}{\rm Ca}$ at energy of pion beam of $E_{\pi^{+}} = 116$~MeV used in Fig.~1~(a,b)
(see Table~2, p.~760 in Ref.~\cite{Akhter.2001.JPG})
\begin{equation}
\begin{array}{llll}
  U = 24.15\, {\rm MeV}, & R_{u} = 1.50\, {\rm fm}, & a_{u} = 0.20\, {\rm fm}.
\end{array}
\label{eq.add.1}
\end{equation}
In calculations in Fig.~\ref{fig.1}~(a) we fix these parameters $R_{u}$, $a_{u}$ and vary $U$ for the different spectra.
Case of $U=24.15$~MeV is presented by upper spectrum (see solid blue line in this figure).
The same spectrum for calculations by optical model is presented in Fig.~\ref{fig.1}~(b)
[see red dashed line in this figure, parameters are in Eq.~(\ref{eq.add.1})].

\begin{figure}[htbp]
\centerline{\includegraphics[width=85mm]{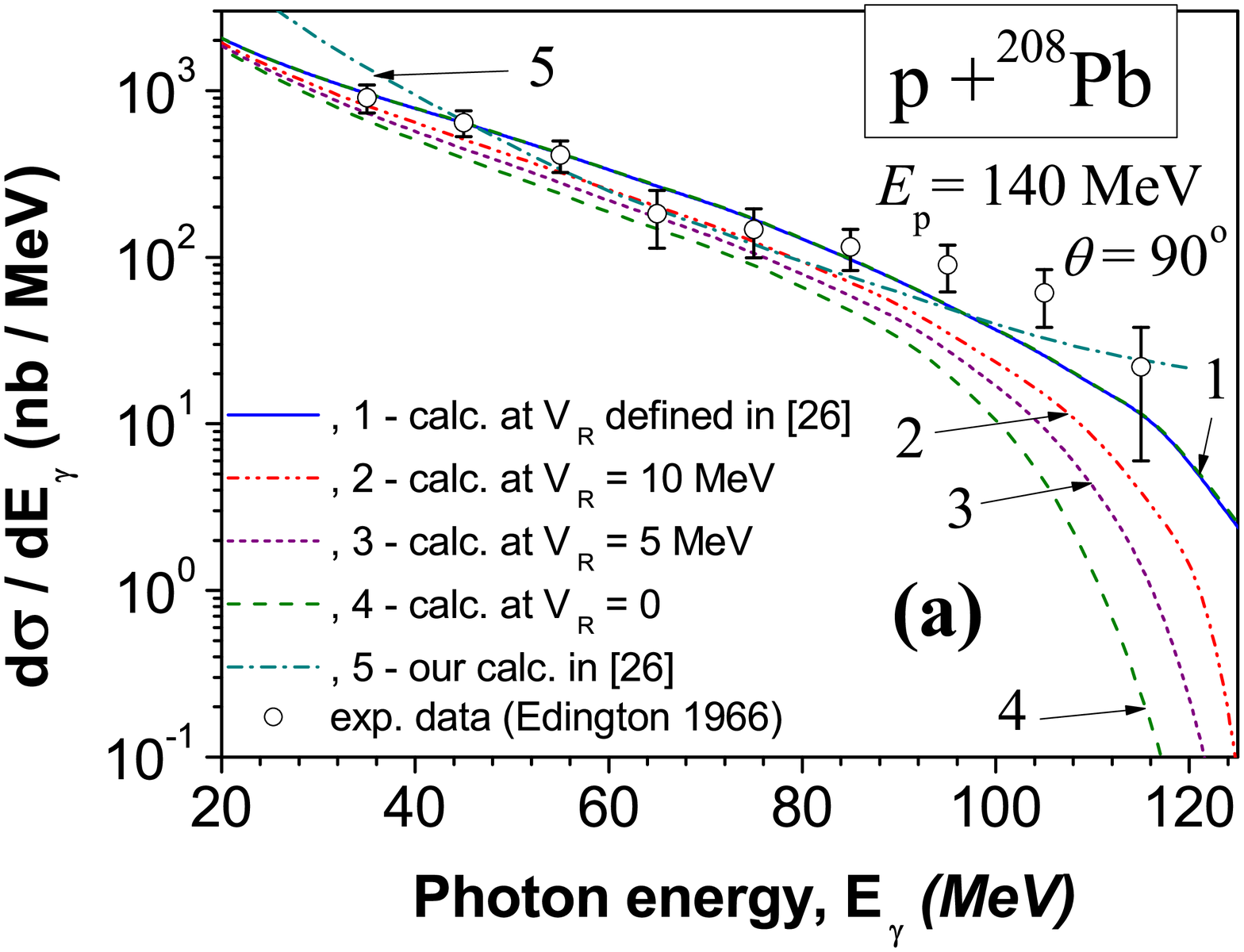}
\hspace{-1mm}\includegraphics[width=85mm]{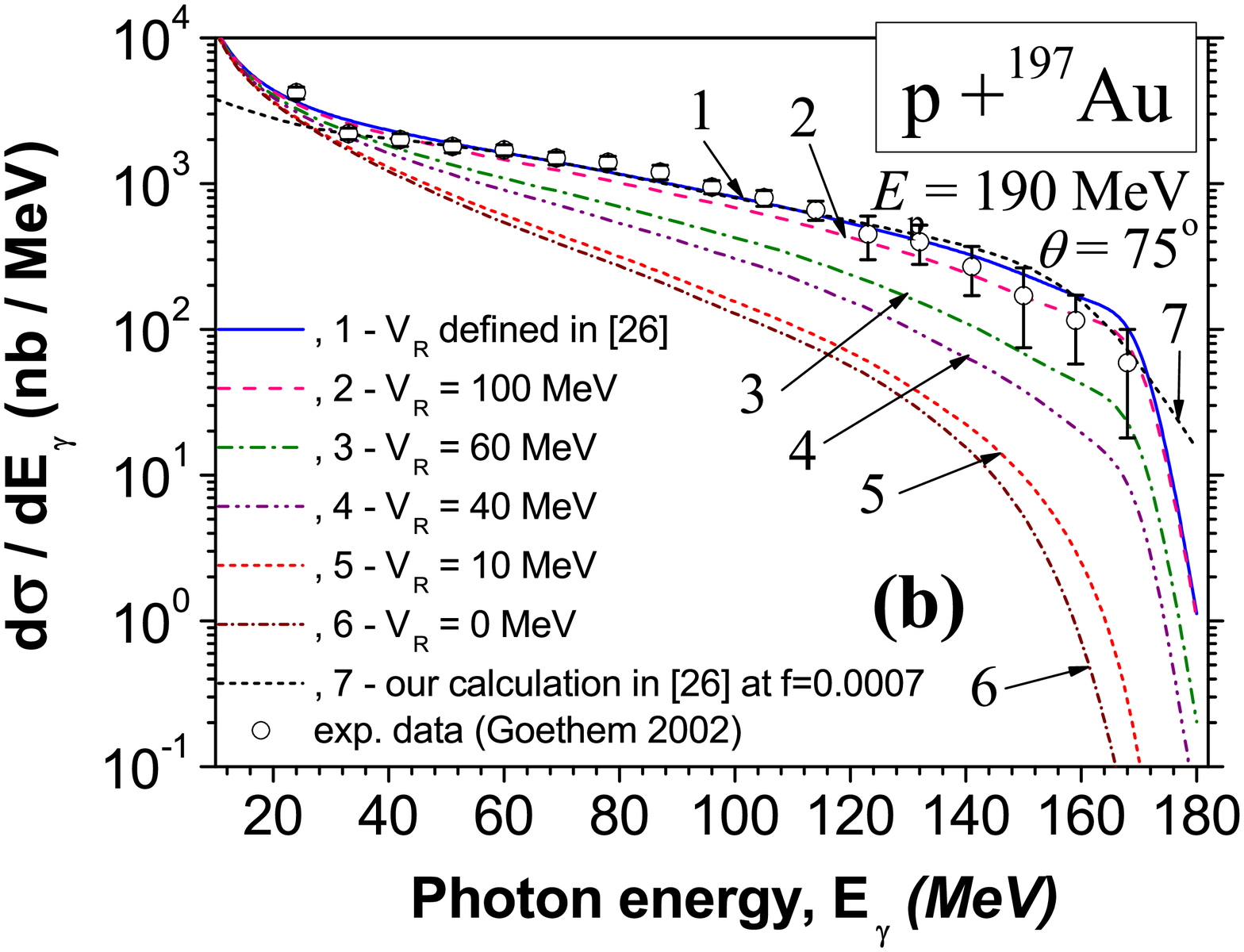}}
\vspace{-4mm}
\caption{\small (Color online)
The bremsstrahlung cross sections of the emitted photons during the scattering of
the $p + ^{208}{\rm Pb}$ at the proton bombarding energy $E_{\rm p}=140$~MeV (a) and $p + ^{197}{\rm Au}$ at the proton bombarding energy $E_{\rm p}=190$~MeV (b)
in comparison with experimental data  (Edington 1966:~\cite{Edington.1966.NP}) 
and (Goethem 2002:~\cite{Goethem.2002.PRL}) 
(circular points)
[
calculations are along formalism~\cite{Maydanyuk.2012.PRC} with electric component $p_{\rm el}$ defined in Eqs.~(36) in that paper
(without magnetic terms $p_{\rm mag, 1}$ and $p_{\rm mag, 2}$ in Eqs.~(36) in Ref.~\cite{Maydanyuk.2012.PRC},
without incoherent bremsstrahlung studied in Ref.~\cite{Maydanyuk_Zhang.2015.PRC}),
parameters of proton-nucleus potential are taken in Eqs.~(46)--(47) in Ref.~\cite{Maydanyuk_Zhang.2015.PRC}
with approximation of $r_{c} = r_{R} = 0.95$~fm].
One can see change of the spectra at high energy region in dependence on nuclear strength $V_{R}$ of the proton-nucleus potential.
\label{fig.4}}
\end{figure}

As we noted, for bremsstrahlung in the proton-nucleus scattering, the understanding of influence of nuclear interactions on the spectrum of photons has not been obtained~\cite{Kopytin.1997.YF}.
Leading contribution to the spectrum in this reaction can be estimated via optical model calculations (where we obtain coherent photons).
Similarity in the bremsstrahlung calculations based on optical model between the proton-nucleus scattering and the pion-nucleus scattering can be found.
Thus, it could be useful to clarify if there is similar sensitivity of the bremsstrahlung spectrum (at high energies of photons)
on nuclear potential for the proton-nucleus scattering.
We 
observe stable dependence of the bremsstrahlung spectra on nuclear strength $V_{R}$ of the proton-nucleus potential at high energy region (see Fig.~\ref{fig.4}).
Note that clear such a dependence of the spectra on the nuclear parameters of proton-nucleus potential has never been found previously.

%

In the proton-nucleus scattering we define the potential along formalism in Ref.~\cite{Maydanyuk_Zhang.2015.PRC} (see Eqs.~(46)--(47) in that paper; but here we follow the detailed study in Ref.~\cite{Becchetti.1969.PR}, which is highly cited and has been tested by many people).
In calculation of the bremsstrahlung spectra in Fig.~\ref{fig.4}~(a,b), we vary the nuclear strength $V_{R}$.
The best agreement between calculations and experimental data corresponds to values of $V_{R}$, defined along formalism in Ref.~\cite{Maydanyuk_Zhang.2015.PRC}
(for both reactions).
In this research we focus on clarifying not small difference between spectra at varying $V_{R}$.
The clear understanding of dependence of the spectra on the nuclear parameter can be obtained, if the incoherent bremsstrahlung is omitted.
Thus, we do not use such terms in the current calculations.
We find that it is enough to establish not small dependence of the spectra on the parameter $V_{R}$.
But, it is not enough to conclude about realistic value for this parameter,
as it would be better to include terms of the incoherent emission in analysis
(that can change the spectra at low energies of photons).
There is also spin-orbital term in the proton-nucleus potential (see $v_{\rm so} (r)$ in Eqs.~(46) in Ref.~\cite{Maydanyuk_Zhang.2015.PRC}; in contrast to the pion-nucleus scattering).
But, we estimate the small bremsstrahlung emission formed by such a term.


\section{Perspectives to study incoherent bremsstrahlung
\label{sec.discussions.spins}}

In Eq.~(\ref{eq.2.3.2}) spins of the nucleons of nucleus are not included.
We studied the role of spins of individual nucleons of full evolving nuclear system in the formation of bremsstrahlung emission during scattering of protons off nuclei~\cite{Maydanyuk.2012.PRC,Maydanyuk_Zhang.2015.PRC}, and during $\alpha$ decay~\cite{Maydanyuk_Zhang_Zou.2016.PRC}.
For the proton-nucleus scattering, the full bremsstrahlung emission can be separated on incoherent bremsstrahlung (i.e. bremsstrahlung formed due to interaction of the incoming proton with the nucleus as a whole),
and coherent bremsstrahlung (i.e. bremsstrahlung formed due to interaction of the incoming proton with the individual nucleons in the nucleus)
[we will omit possible interference terms].

In previous study~\cite{Maydanyuk.2012.PRC,Maydanyuk_Zhang.2015.PRC,Maydanyuk_Zhang_Zou.2016.PRC} we took into account both aspects.
The full matrix element of emission has term corresponding to the incoherent emission and terms corresponding to different aspects of the coherent emission.
This allows us to estimate contributions of each component to the full bremsstrahlung spectrum. That was realized for the proton-nucleus scattering and $\alpha$ decay (partially), but not for pion-nucleus scattering.
However, in the current research we would like to restrict ourselves by study of the coherent emission because of the following reasons.

\begin{enumerate}
\item
Study of the incoherent bremsstrahlung is a more complicated task, from mathematical and numerical point of view.
We do not find any information about emission of bremsstrahlung photons during pion-nucleus scattering in the literature.
Thus, our calculations with results of other people cannot be compared (we study bremsstrahlung in pion-nucleus scattering at first time).
We estimate that the incoherent emission (its contribution to the full spectrum) in the pion-nucleus scattering is essentially sensitive basing on the coherent bremsstrahlung calculations.
Because of this, we want initially to obtain proper basis for description (and estimation) of coherent bremsstrahlung in the pion nucleus scattering.
Thus, we put a main focus on construction of model of the coherent bremsstrahlung in the pion-nucleus scattering.

\item
Study of spins of nucleons and role of individual nucleons of the nucleus is included in our interest of study in this topic.
We plan to do it in this research area, as a natural next step.
We plan to begin with formalism~\cite{Maydanyuk_Zhang.2015.PRC} [see Sect. E, F, Eqs.~(33--37, 41, 43--45) in that paper, which includes spin terms]
with needed generalization.

\item
If to return back to our previous study of role of individual nucleons of the nucleus and spin of scattered proton, then one can find the following.
The experimental bremsstrahlung spectra \cite{Goethem.2002.PRL,Clayton.1992.PRC,Clayton.1991.PhD} in the proton-nucleus scattering have the hump-shaped plateau inside the middle energy region. And they decrease to the kinematic high energy limit of the photons
(these are data for the scattering of $p + ^{208}{\rm Pb}$ at the proton beam energies of 140 and 145~MeV,
the scattering of $p + ^{12}{\rm C}$, $p + ^{58}{\rm Ni}$, $p + ^{107}{\rm Ag}$ and $p + ^{197}{\rm Au}$ at the proton beam energy of 190~MeV).
However, the experimental bremsstrahlung spectra~\cite{Boie.2007.PRL,Boie.2009.PhD,Maydanyuk.2008.EPJA,Giardina.2008.MPLA}
in the $\alpha$ decay of the $^{210, 214}{\rm Po}$, $^{226}{\rm Ra}$ nuclei have shape of the logarithmic type without humped-shaped such a form.
Using unified formalism~\cite{Maydanyuk.2012.PRC,Maydanyuk_Zhang.2015.PRC,Maydanyuk_Zhang_Zou.2016.PRC},
we explained this difference between the bremsstrahlung spectra in the proton--nucleus scattering and the $\alpha$ decay.
In the proton-nucleus scattering, non-zero spin of the scattered proton gives a new type of interaction, based on its relations with momenta of nucleons of the nucleus [see $p_{4}$ in (43) and (44), in Ref.~\cite{Maydanyuk_Zhang.2015.PRC}]. This produces the incoherent type of the bremsstrahlung emission, which is absent in the $\alpha$ decay.
Inclusion of the matrix element of incoherent emission $p_{4}$ for proton-nucleus scattering to calculations changes the bremsstrahlung spectra (humped-shaped plateau appears in the spectra) and allows to describe experimental data well.

According to results of Refs.~\cite{Maydanyuk_Zhang_Zou.2016.PRC}, at higher energies of photons the internal structure of the $\alpha$-particle in the $\alpha$ decay has a more important role in interactions between the $\alpha$-particles and nuclei.
This would produce incoherent contribution to the full bremsstrahlung spectrum.

So, if to consider logic above and take zero spin for pion into account, then the small incoherent bremsstrahlung emission in pion-nucleus scattering can be established.
However, this type of emission is increased at higher energies of photons and pions.
But, this task is a more complicated, and it is sensible to focus on it in further research (after obtaining understanding of coherent emission in pion-nucleus scattering as a proper basis).
\end{enumerate}

\section{Comparison of our model with
approach~\cite{Gil.1998.PLB}
\label{sec.discussions.comparisonmodelOset}}

As we noted above (see Sect.~\ref{sec.introduction}), in Ref.~\cite{Gil.1998.PLB} emission of the coherent photons by virtual pions was studied,
which are produced in the nuclear matter during the proton - nucleus scattering.
However, we find that this type of emission is essentially different from photons studied in our work.
In formation of photons during reaction studied in Ref.~\cite{Gil.1998.PLB} the Coulomb forces outside the nucleus-target are not included in consideration.
But term of the bremsstrahlung photons emitted by such Coulomb forces outside the nucleus gives the largest contribution to the full coherent and incoherent
bremsstrahlung emission in the scattering of real pions (in beams) off nuclei.
If to include such a contribution in calculations of full bremsstrahlung spectrum, then the role of the nuclear forces will be essentially suppressed (i.e. it will be essentially more difficult to estimate the role of nuclear interactions via analysis of the bremsstrahlung spectra).
But, in our paper we find that this is possible also.

Comparing our formalism with the paper~\cite{Gil.1998.PLB}, we find the following:

\begin{itemize}
\item
We base our calculations of the matrix elements on the wave functions of pion-nucleus system, along to main positions of quantum mechanics.
These wave functions are complex, continuous in the full space region of definition. They are defined for the boundary conditions chosen for the studied reaction.
Because of this, we take into account possible interference effects in calculations of bremsstrahlung cross-sections
(in contrast to approach~\cite{Gil.1998.PLB}).
This shall become important, if we study contributions of different types of emission of photons to the full bremsstrahlung spectrum,
and estimate nuclear interactions from analysis of experimental bremsstrahlung data.

\item
We obtain direct correspondence between interactions for different nuclei and wave functions.
In some cases, these wave functions can be essentially different for
(1) different nuclei or different isotopes in the scattering,
(3) the same nuclei in the scattering (in dependence on the chosen internal mechanisms inside compound nuclear system)
[i.e. description of nuclear matter can be essentially different for bound states of nuclei without the scattering, and for unbound states of nuclei in the scattering],
etc..

\item
For example, in Ref.~\cite{Maydanyuk.2015.NPA,Maydanyuk_Zhang_Zou.2017.PRC} we show that a more accurate stationary consideration of formation of compound nuclear system during the scattering (even without inclusion of inelastic mechanisms) can change cross-sections essentially
(up to 4 times, i.e. more than 100 percents, for reactions $\alpha + ^{40,44,48}{\rm Ca}$),
even for the same full wave functions with the same boundary conditions.
Reason is in more accurate description of internal mechanisms inside the full nuclear system, which can be characterized via different penetrabilities (for the same wave functions).
Such quantum effects exists also in the scattering of protons and pins off nuclei.
Approach~\cite{Gil.1998.PLB} does not take not small such an effect into account (our formalism allows to include description of such effects completely).

\item
Our formalism is based on Hartree approximation in determination of the full wave function (of pion-nucleus and proton-nucleus systems).
Anti-symmetrization of one-nucleon wave functions in construction of the wave function of the nucleus is used in a more complete way (that reflects Pauli exclusion principle for nucleons). This aspect can be taken into account in our description of emission of photons.

\end{itemize}

Thus, our model uses a more complete basis in quantum description of the scattering, in comparison with formalism~\cite{Gil.1998.PLB}.
This way allows to describe the scattering process accurately, basing on experimental information which we know and which is extracted on the basis of theory of
nuclear reactions.
If we want to extract information about interactions between the nuclei and pions (or protons) from the bremsstrahlung measurements, this aspect becomes a more important.
But, in the current research we do not analyze resonances studied in Ref.~\cite{Gil.1998.PLB}.
This question is subject of perspective inclusion to further research.

\section{Conclusions 
\label{sec.conclusions}}

In this paper we have performed the first investigations of emission of the bremsstrahlung photons during the scattering of pions off nuclei.
A motivation of this research is our supposition that such photons can
be used as a new independent probe of non-Coulomb part of the pion-nucleus interactions.
We construct a new model of the bremsstrahlung photons emitted in this reaction.
To describe the interactions between the $\pi^{\pm}$-mesons and nuclei, we use two nuclear potentials:
(1) the Kisslinger-type potential along Johnson-Satchler formalism obtained by the Krell-Ericson transformation from Klein-Gordon equation for the pion scattering~\cite{Johnson.1996.AP}, and
(2) the Woods-Saxon potential used by Akhter et al. in optical model calculations~\cite{Akhter.2001.JPG}.
We find that emission of photons formed due to nuclear part of the Johnon-Satchler potential gives essential contribution to the full spectrum.
Thus, according to formalism~\cite{Johnson.1996.AP}, nuclear interactions play an important (large) role in the formation of the bremsstrahlung spectra.
Moreover, they can be studied experimentally via measurements of the bremsstrahlung photons.

Importance of such a result is reinforced 
if to remind that,
in exception with paper~\cite{Maydanyuk_Zhang_Zou.2016.PRC},
it has never been possible to extract any information about nuclear parameters of optical model in nuclear reactions from
analysis of existing experimental data of the accompanying bremsstrahlung.
This is because the Coulomb interactions play a larger role in the forming the bremsstrahlung, than the nuclear interactions.
%
Our results in pion-nucleus scattering show that bremsstrahlung spectra are essentially sensitive on non-Coulomb and nuclear parameters in high energy region of photons.
For the first time, we observe similar dependence of the spectra on the strength of nuclear optical potential for proton-nucleus scattering (see Fig.~\ref{fig.4}).
Thus, possible measurements of the bremsstrahlung photons could be a good tools for obtaining a new information about the non-Coulomb and nuclear interactions between $\pi$-mesons and nuclei.

\section*{Acknowledgements
\label{sec.acknowledgements}}

Authors are grateful to Prof. Rolf H. Siemssen and Prof. Hans W. Wilschut for interesting discussions concerning opportunities of experiments with $\pi$-mesons.
S.~P.~Maydanyuk thanks the Institute of Modern Physics of Chinese Academy of Sciences for warm hospitality and support.
This work was supported by the Major State Basic Research Development Program in China (No. 2015CB856903),
the National Natural Science Foundation of China (Grant Nos. 11575254, 11447105 and 11175215),
the Chinese Academy of Sciences fellowships for researchers from developing countries (No. 2014FFJA0003).


\appendix
\section{Operator of emission for evolving many-nucleon system and $\pi^{\pm}$-meson
\label{sec.app.1}}


Emission of the bremsstrahlung photons can be introduced to the formalism of the $\pi^{\pm}$-nucleus scattering via application of Coulomb gauge for each electromagnetic charge in the system
as
\begin{equation}
\begin{array}{lcl}
   \mathbf{p}_{i} \to \mathbf{p}_{i} - \displaystyle\frac{z_{i}\, e}{c}\, \mathbf{A} (\mathbf{r}_{i}, t),
\end{array}
\label{eq.app.2.0.1}
\end{equation}
where
$\mathbf{p}_{i} = - i\hbar\, \nabla_{i}$ is momentum of pion or nucleon with number $i$,
$\mathbf{A} (\mathbf{r}_{i}, t)$ is the vector potential of the electromagnetic field formed by motion of pion or nucleon with number $i$,
$z_{i}$ is the electromagnetic charge of pion or nucleon with number $i$.
The modified Hamiltonian is written as $\hat{H} = \hat{H}_{0} + \hat{H}_{\gamma}$,
where $\hat{H}_{\gamma}$ is a new correction (i.e. operator of emission) describing emission of the bremsstrahlung photons.

We begin with generalization of the Pauli equation on the system composed of $A$ nucleons and $\pi^{\pm}$,
describing scattering of $\pi^{\pm}$ off nucleus with $A$ nucleons
with Hamiltonian constructed as
(see Eqs.~(1)--(2) in Ref.~\cite{Maydanyuk_Zhang.2015.PRC}, also see Ref.~\cite{Maydanyuk.2012.PRC} for the proton--nucleus scattering)
\begin{equation}
\begin{array}{lcl}
  \hat{H} =
  \displaystyle\sum_{i=1}^{A+1}
  \biggl\{
    \displaystyle\frac{1}{2\,m_{i}}\;
    \Bigl( \mathbf{p}_{i} - \displaystyle\frac{z_{i}e}{c} \mathbf{A}_{i} \Bigr)^{2} -
    \displaystyle\frac{z_{i}e\hbar}{2m_{i}c}\, \sigmabf \cdot \mathbf{rot A}_{i} +
    z_{i}e\, A_{i,0}
  \biggr\} +
  V(\mathbf{r}_{1} \ldots \mathbf{r}_{A+1}) = \hat{H}_{0} + \hat{H}_{\gamma},
\end{array}
\label{eq.app.2.1.1}
\end{equation}
where
\begin{equation}
\begin{array}{lcl}
  \hat{H}_{0} =
  \displaystyle\sum_{i=1}^{A+1}
    \displaystyle\frac{1}{2\,m_{i}}\: \mathbf{p}_{i}^{2} +
  V(\mathbf{r}_{1} \ldots \mathbf{r}_{A+1}), \\
  \hat{H}_{\gamma} =
  \displaystyle\sum_{i=1}^{A+1}
  \biggl\{
    - \displaystyle\frac{z_{i} e}{m_{i}c}\; \mathbf{p}_{i} \mathbf{A}_{i} +
    \displaystyle\frac{z_{i}^{2}e^{2}}{2m_{i}c^{2}} \mathbf{A}_{i}^{2} -
    \displaystyle\frac{z_{i}e\hbar}{2m_{i}c}\, \sigmabf \cdot \mathbf{rot A}_{i} +
    z_{i}e\, A_{i,0}
  \biggr\}.
\end{array}
\label{eq.app.2.1.2}
\end{equation}
Here, $m_{i}$ and $z_{i}$ are mass and electromagnetic charge of nucleon with number $i$ ($i$=1 \ldots A)
or mass and electromagnetic charge of $\pi^{\pm}$-meson ($i=A+1$),
$\mathbf{p}_{i} = -i\hbar\, \mathbf{d}/\mathbf{dr}_{i} $ is momentum operator for nucleon with number $i$ ($i$=1 \ldots A) or $\pi^{\pm}$-meson ($i=A+1$),
$V(\mathbf{r}_{1} \ldots \mathbf{r}_{A+1})$ is general form of the potential of interactions between nucleons and $\pi^{\pm}$-meson,
$\sigmabf$ are Pauli matrixes,
$A_{i} = (\mathbf{A}_{i}, A_{i,0})$ is potential of electromagnetic field formed by moving nucleon with number $i$ ($i$=1 \ldots A) or $\pi^{\pm}$-meson ($i=A+1$).
Let us turn to the center of masses frame.
Introducing coordinates of centers of masses for the nucleus
$\mathbf{R}_{A} = \sum_{j=1}^{A} m_{j}\, \mathbf{r}_{A j} / m_{A}$,
coordinates of centers of masses of the complete system
$\mathbf{R} = (m_{A}\mathbf{R}_{A} + m_{\pi}\mathbf{r}_{\pi}) / (m_{A}+m_{\pi})$,
relative coordinates $\rhobf_{A j} = \mathbf{r}_{j} - \mathbf{R}_{A}$ and
$\mathbf{r} = \mathbf{r}_{\pi} - \mathbf{R}_{A}$,
we obtain new independent variables $\mathbf{R}$, $\mathbf{r}$ and
$\rhobf_{Aj}$ ($j=1 \ldots A-1$)%
\footnote{Sometimes, we will add additional bottom index $A$ to variables of nucleons, for better clearness.}
\begin{equation}
\begin{array}{lll}
  \mathbf{R} =
    \displaystyle\frac{1}{m_{A}+m_{\pi}}\;
    \Bigl\{
      \displaystyle\sum_{j=1}^{A} m_{Aj}\, \mathbf{r}_{A j} +
      m_{\pi}\, \mathbf{r}_{\pi}
    \Bigr\}, &
  \mathbf{r} =
  \mathbf{r}_{\pi} - \displaystyle\frac{1}{m_{A}} \displaystyle\sum_{j=1}^{A}
  m_{Aj}\, \mathbf{r}_{Aj}, &
  \rhobf_{A j} =
  \mathbf{r}_{A j} -
  \displaystyle\frac{1}{m_{A}} \displaystyle\sum_{k=1}^{A} m_{Ak}\, \mathbf{r}_{Ak},
\end{array}
\label{eq.app.2.1.3}
\end{equation}
and calculate operators of corresponding momenta
\begin{equation}
\begin{array}{lclll}
  \mathbf{p}_{\pi} =
    - i\hbar\, \displaystyle\frac{\mathbf{d}}{\mathbf{dr}_{\pi}} =
    \displaystyle\frac{m_{\pi}}{m_{A} + m_{\pi}}\, \mathbf{P} + \mathbf{p}, &
  \mathbf{p}_{Aj} =
    - i\hbar\, \displaystyle\frac{\mathbf{d}}{\mathbf{dr}_{Aj}} =
    \displaystyle\frac{m_{Aj}}{m_{A} + m_{\pi}}\, \mathbf{P} -
    \displaystyle\frac{m_{Aj}}{m_{A}}\,\mathbf{p} +
    \displaystyle\frac{m_{A} - m_{Aj}}{m_{A}}\; \mathbf{\tilde{p}}_{Aj} -
    \displaystyle\frac{m_{Aj}}{m_{A}}\,
    \displaystyle\sum_{k=1, k \ne j}^{A-1} \mathbf{\tilde{p}}_{A k},
\end{array}
\label{eq.app.2.1.4}
\end{equation}
where
$\mathbf{P} = -i\hbar\, \mathbf{d}/\mathbf{dR}$,
$\mathbf{p} = -i\hbar\, \mathbf{d}/\mathbf{dr}$,
$\mathbf{\tilde{p}}_{Aj} = -i\,\hbar\, \mathbf{d}/\mathbf{d}\rhobf_{Aj}$,
$m_{\pi}$ and $m_{A}$ are masses of the scattering $\pi^{\pm}$ and nucleus.


Let us study leading emission operator of the system composed of $\pi^{\pm}$
and nucleus in the laboratory frame. We obtain its view in Eq.~(\ref{eq.app.2.1.2})
neglecting terms at $\mathbf{A}_{i}^{2}$, $A_{i,0}$ and spinor term:
\begin{equation}
\begin{array}{lcl}
  \hat{H}_{\gamma} =
    - \displaystyle\frac{z_{\pi}\,e}{m_{\pi}c}\;
    \mathbf{A}(\mathbf{r}_{\pi},t)\, \mathbf{\hat{p}}_{\pi} -
    \displaystyle\sum\limits_{j=1}^{A}
    \displaystyle\frac{z_{j}\,e}{m_{j}c}\;
    \mathbf{A}(\mathbf{r}_{j},t)\, \mathbf{\hat{p}}_{j}.
\end{array}
\label{eq.app.2.3.1}
\end{equation}
Here, $\mathbf{A}(\mathbf{r}_{s},t)$ describes emission of photon
caused by $\pi^{\pm}$-meson or nucleon ($s=\pi$ is for $\pi^{\pm}$, $s=j$ for the nucleons of the nucleus).
Using its presentation in the form~(5) of~\cite{Maydanyuk.2012.PRC},
for the emission operator 
in the laboratory frame we obtain:
\begin{equation}
\begin{array}{lcl}
  \hat{H}_{\gamma} =
  -\,e\, \sqrt{\displaystyle\frac{2\pi\hbar}{w_{\rm ph}}}\,
    \displaystyle\sum\limits_{\alpha=1,2} \mathbf{e}^{(\alpha),*}\;
    e^{-i \mathbf{k}\, \bigl[\mathbf{R} - \displaystyle\frac{m_{\pi}}{M+m_{\pi}}\:
    \mathbf{r}\bigr]}\;
  \Biggl\{
    \displaystyle\frac{1}{M + m_{\pi}}\;
    \biggl[
      e^{-i \mathbf{k}\, \mathbf{r}}\; z_{\pi} +
      \displaystyle\sum\limits_{j=1}^{A} z_{Aj}\; e^{-i \mathbf{k}\, \rhobfsm_{Aj}}
    \biggr]\; \mathbf{P}\; +\; \\
  + \;
    \biggl[
      e^{-i \mathbf{k}\, \mathbf{r}}\; \displaystyle\frac{z_{\pi}}{m_{\pi}}\; -
      \displaystyle\frac{1}{M}
      \displaystyle\sum\limits_{j=1}^{A}
        z_{Aj}\; e^{-i \mathbf{k}\, \rhobfsm_{Aj}}
    \biggr]\; \mathbf{p}\; + \;
  \displaystyle\sum\limits_{j=1}^{A-1}
      \displaystyle\frac{z_{Aj}}{m_{Aj}}\; e^{-i \mathbf{k}\,
      \rhobfsm_{Aj}}\, \mathbf{\tilde{p}}_{Aj}\; -\;
  \displaystyle\frac{1}{M}
  \biggl[
    \displaystyle\sum\limits_{j=1}^{A}
      z_{Aj}\; e^{-i \mathbf{k}\, \rhobfsm_{Aj}}
  \biggr]\; \displaystyle\sum_{k=1}^{A-1} \mathbf{\tilde{p}}_{Ak}
  \Biggr\},
\end{array}
\label{eq.app.2.3.2}
\end{equation}
where star denotes complex conjugation,
$\mathbf{e}^{(\alpha)}$ are unit vectors of polarization of the photon emitted
($\mathbf{e}^{(\alpha),*} = \mathbf{e}^{(\alpha)}$), $\mathbf{k}$ is the wave vector of the photon, and
$w_{\rm ph} = k\, c = \bigl| \mathbf{k}\bigr|\, c$.
Vectors $\mathbf{e}^{(\alpha)}$ are perpendicular to $\mathbf{k}$ in the Coulomb gauge.
We have two independent polarizations $\mathbf{e}^{(1)}$ and $\mathbf{e}^{(2)}$
for the photon with momentum $\mathbf{k}$ ($\alpha = 1,2$).
%

\section{Wave function of the $\pi^{\pm}$-nucleons system
\label{sec.app.2}}

Emission of the bremsstrahlung photons is caused by the relative motion of nucleons and the charged $\pi^{\pm}$-meson of the full system. However, we assume that the most intensive emission of photons is formed by relative motion of $\pi^{\pm}$ related to the nucleus. Thus, it is sensible to represent the total wave function via coordinates of relative motion of these complicated objects.
Following to such a logic, we define the wave function of the full nuclear system as
\begin{equation}
  \Psi_{s} =
  \Phi_{s} (\mathbf{R}) \cdot
  \Phi_{\rm \pi -nucl, s}(\mathbf{r}) \cdot
  \psi_{\rm nucl, s}(\beta_{A}),
\label{eq.app.2.4.1}
\end{equation}
where
$s = i$ or $f$ (indexes $i$ and $f$ denote the initial state, i.e. the state before emission of photon,
and the final state, i.e. the state after emission of photon),
$\mathbf{K}_{s}$ is full momentum of the $\pi^{\pm}$--nucleus system (in the laboratory frame),
$\Phi_{s} (\mathbf{R})$ is wave function describing motion of center-of-masses of the full nuclear system in the laboratory frame,
$\Phi_{\rm \pi-nucl,s} (\mathbf{r})$ is function describing relative motion (with tunneling for under-barrier energies) of $\pi^{\pm}$ related to the nucleus (without description of internal relative motions of nucleons inside the nucleus),
$\psi_{\rm nucl,s}(\beta)$ is the many-nucleon function describing internal states of nucleons in the nucleus
(it determines space states on the basis of relative distances $\rhobf_{1}$ \ldots $\rhobf_{A}$ of nucleons of the nucleus related to its center-of-masses, and spin-isospin states also),
$\beta_{A}$ is a set of numbers $1 \ldots A$ of nucleons of the nucleus.
We assume:
\begin{equation}
\begin{array}{lcl}
  \Phi_{\bar{s}} (\mathbf{R}) =  N_{\bar{s}}\; e^{-i\,\mathbf{K}_{\bar{s}}\cdot\mathbf{R}}, & \mathbf{K}_{i} = 0,
\end{array}
\label{eq.app.2.4.2}
\end{equation}
where $N_{\bar{s}}$ is normalized factor which will be defined later.

Motion of nucleons of the nucleus relative to each other does not have large influence on the states describing relative motion of $\pi^{\pm}$ related to the nucleus. Therefore, such a representation of the full wave function can be considered as an approximation.
However, the relative internal motions of nucleons of the nucleus give own contributions to the full bremsstrahlung spectrum and they can be estimated.
We include the many-nucleon structure in wave function $\psi_{\rm nucl, s}(\beta_{A})$ of the nucleus while we assume that wave function of relative motion $\psi_{\rm \pi-nucl,s}(\mathbf{r})$ is calculated without it but with maximal orientation of the $\pi^{\pm}$-nucleus potential well extracted from experimental data of $\pi^{\pm}$-nucleus scattering (one can take the many-nucleon corrections into account in the next step). Such a line allows us to keep accurately the wave function of relative motion which gives the leading contribution to the bremsstrahlung spectrum,
while the many-nucleon structure should be estimated after as a correction.
\section{Matrix element of emission
\label{sec.app.3}}

We find the matrix element on the basis of the operator of emission (\ref{eq.app.2.3.2}) and
wave function (\ref{eq.app.2.4.1}):
\begin{equation}
\begin{array}{lcl}
  \vspace{0mm}
  \langle \Psi_{f} |\, \hat{H}_{\gamma} |\, \Psi{i} \rangle & = &

  -\, N_{i}\, N_{f}\; e\;
  \sqrt{\displaystyle\frac{2\pi\hbar}{w_{\rm ph}}}\,

    \displaystyle\sum\limits_{\alpha=1,2} \mathbf{e}^{(\alpha),*}\;
  \Biggl\{ \\

  \vspace{2mm}
  & & \times \;
  \Biggl\langle \Psi_{f} \Biggl|\,
    e^{i\,(\mathbf{K}_{i} - \mathbf{K}_{f} - \mathbf{k})\cdot\mathbf{R}}\:
    e^{i\, \mathbf{kr}\, \displaystyle\frac{m_{\pi}}{m_{A}+m_{\pi}}}\;
    \displaystyle\frac{1}{m_{A} + m_{\pi}}\;
    \biggl[
      e^{-i \mathbf{k}\, \mathbf{r}}\, z_{\pi} +
      \displaystyle\sum\limits_{j=1}^{A} z_{Aj}\; e^{-i \mathbf{k}\, \rhobfsm_{Aj}}
    \biggr]\; \mathbf{P}\;
    \Biggr|\, \Psi_{i} \Biggr\rangle\; + \\

  \vspace{2mm}
  & & +\;
    \Biggl\langle \Psi_{f} \Biggl|\,
      e^{i\,(\mathbf{K}_{i} - \mathbf{K}_{f} - \mathbf{k})\cdot\mathbf{R}}\:
      e^{i\, \mathbf{kr}\, \displaystyle\frac{m_{\pi}}{m_{A}+m_{\pi}}}\,
    \biggl[
      e^{-i \mathbf{k}\, \mathbf{r}}\, \displaystyle\frac{z_{\pi}}{m_{\pi}} -
      \displaystyle\sum\limits_{j=1}^{A}
        \displaystyle\frac{z_{Aj}}{m_{A}}\; e^{-i \mathbf{k}\, \rhobfsm_{Aj}}
    \biggr]\; \mathbf{p}\;
    \Biggr|\, \Psi_{i} \Biggr\rangle\; + \\

  \vspace{2mm}
  & & +\;
    \Biggl\langle \Psi_{f} \Biggl|\,
      e^{i\,(\mathbf{K}_{i} - \mathbf{K}_{f} - \mathbf{k})\cdot\mathbf{R}}\:
      e^{i\, \mathbf{kr}\, \displaystyle\frac{m_{\pi}}{m_{A}+m_{\pi}}}\,
      \biggl[
        \displaystyle\sum\limits_{j=1}^{A-1}
          \displaystyle\frac{z_{A j}}{m_{Aj}}\:
          e^{-i \mathbf{k}\, \rhobfsm_{Aj}}\,
          \mathbf{\tilde{p}}_{Aj}
      \biggr]\; \Biggr|\, \Psi_{i} \Biggr\rangle\; - \\

  \vspace{2mm}
  & & -\;
    \Biggl\langle \Psi_{f} \Biggl|\,
      e^{i\,(\mathbf{K}_{i} - \mathbf{K}_{f} - \mathbf{k})\cdot\mathbf{R}}\:
      e^{i\, \mathbf{kr}\, \displaystyle\frac{m_{\pi}}{M+m_{\pi}}}\,
    \displaystyle\frac{1}{m_{A}}\,
    \biggl[
      \displaystyle\sum\limits_{j=1}^{A}
      z_{Aj}\; e^{-i \mathbf{k}\, \rhobfsm_{Aj}}\;
      \displaystyle\sum_{k=1}^{A-1} \mathbf{\tilde{p}}_{Ak}
    \biggr]\; \Biggr|\, \Psi_{i} \Biggr\rangle\;
  \Biggr\}.
\end{array}
\label{eq.app.2.5.1}
\end{equation}
The first term describes emission of photon caused by motion of the full nuclear system in the laboratory frame and its response on the emission of photon. We shall calculate the spectra in the center-of-mass frame, so we shall neglect this term.
The second term describes emission of photon caused by $\pi^{\pm}$-meson and each nucleon of the nucleus, at relative motion of $\pi^{\pm}$ concerned with the nucleus. This term is leading and gives main contribution to the full bremsstrahlung spectrum.
The third and fourth terms describe emission of photon caused by each nucleon of the nucleus, in relative motions of nucleons of the nucleus inside its space region
(any nuclear deformations during emission can be connected with such terms).

We consider the leading matrix element on the basis of the second term
in Eq.~(\ref{eq.app.2.5.1}). In calculations, we must integrate over
all independent space variables given in (\ref{eq.app.2.1.3}):
\begin{equation}
\begin{array}{ll}
\vspace{1mm}
  & \langle \Psi_{f} |\, \hat{H}_{\gamma} |\, \Psi_{i} \rangle_{1} \; =\;
  -\, N_{i}\, N_{f}\; e\;
  \sqrt{\displaystyle\frac{2\pi\hbar}{w_{\rm ph}}}\,
    \displaystyle\sum\limits_{\alpha=1,2} \mathbf{e}^{(\alpha),*}\;
    \displaystyle\int
      e^{i\,(\mathbf{K}_{i} - \mathbf{K}_{f} - \mathbf{k})\,\mathbf{R}}\: \mathbf{dR}\; \times \\

  \times &
    \Biggl\langle
      \Phi_{\rm \pi-nucl, f}(\mathbf{r}) \cdot
      \psi_{\rm nucl, f}(\beta_{A})
    \Biggl|\,
      e^{i\, \mathbf{kr}\, \displaystyle\frac{m_{\pi}}{m_{A}+m_{\pi}}}\,
    \biggl[
      e^{-i \mathbf{k}\, \mathbf{r}}\;
        \displaystyle\frac{z_{\pi}}{m_{\pi}} -
      \displaystyle\sum\limits_{j=1}^{A}
        \displaystyle\frac{z_{Aj}}{m_{A}}\; e^{-i \mathbf{k}\, \rhobfsm_{Aj}}
    \biggr]\; \mathbf{p}\;
    \Biggr|\,
      \Phi_{\rm \pi-nucl, i}(\mathbf{r}) \cdot
      \psi_{\rm nucl, i}(\beta_{A})
    \Biggr\rangle.
\end{array}
\label{eq.app.2.5.2}
\end{equation}
Introducing the \emph{effective charge of the $\pi^{\pm}$--nucleus system} as
\begin{equation}
\begin{array}{lcl}
  Z_{\rm eff}(\mathbf{k}, \mathbf{r}) & = &
    e^{i\,\mathbf{kr}\, \displaystyle\frac{m_{\pi}}{m_{A}+m_{\pi}}}\:
    \biggl\{
      \displaystyle\frac{m_{A}\, z_{\pi}}{m_{A}+m_{\pi}} -
      e^{i\,\mathbf{kr}}\:
      \displaystyle\frac{m_{\pi}\, Z_{\rm A}(\mathbf{k})}{m_{A}+m_{\pi}}
    \biggr\}
\end{array}
\label{eq.app.2.5.3}
\end{equation}
and the \definition{charged form-factor of the nucleus} as
\begin{equation}
  Z_{\rm A} (\mathbf{k}) =
  \Bigl\langle \psi_{\rm nucl, f} (\rhobf_{A1} \ldots \rhobf_{AA})\: \Bigl|\;
    \displaystyle\sum\limits_{j=1}^{A}
      z_{Aj}\:
      e^{-i \mathbf{k} \rhobfsm_{Aj}}\:
  \Bigr|\: \psi_{\rm nucl, i} (\rhobf_{A1} \ldots \rhobf_{AA})\, \Bigr\rangle ,
\label{eq.app.2.5.4}
\end{equation}
we obtain:
\begin{equation}
\begin{array}{lcl}
  \langle \Psi_{f} |\, \hat{H}_{\gamma} |\, \Psi_{i} \rangle_{1} =
  -\, N_{i}\, N_{f}\;
  \displaystyle\frac{e}{m}\:
  \sqrt{\displaystyle\frac{2\pi\hbar}{w_{\rm ph}}}\;
  (2\pi)^{3}\:
    \displaystyle\sum\limits_{\alpha=1,2} \mathbf{e}^{(\alpha),*}\;
    \delta(\mathbf{K}_{f} - \mathbf{K}_{i} - \mathbf{k}) \cdot
    \biggl\langle\: \Phi_{\rm \pi-nucl, f} (\mathbf{r})\: \biggl|\,
      Z_{\rm eff}(\mathbf{k}, \mathbf{r})\: e^{-i\,\mathbf{kr}}\: \mathbf{p}\;
    \biggr|\: \Phi_{\rm \pi-nucl, i} (\mathbf{r})\: \biggr\rangle,
\end{array}
\label{eq.app.2.5.5}
\end{equation}
where $m = m_{\pi} m_{A} / (m_{\pi} + m_{A})$ and
we use integral representation of the delta-function as
\begin{equation}
  \displaystyle\int
    e^{i\,(\mathbf{K}_{i} - \mathbf{K}_{f} - \mathbf{k})\,\mathbf{R}}\: \mathbf{dR} =
  (2\pi)^{3}\: \delta (\mathbf{K}_{i} - \mathbf{K}_{f} - \mathbf{k}).
\label{eq.app.2.5.6}
\end{equation}
We define the normalizing factors $N_{i}$ and $N_{f}$ as
\begin{equation}
  N_{i} = N_{f} = (2\pi)^{-3/2}.
\label{eq.app.2.5.7}
\end{equation}
We calculate cross sections of the emitted photons not dependent on momentum $\mathbf{K}_{f}$ (momentum of the full $\pi^{\pm}$-nucleus system after emission of photon in the laboratory frame). Thus, we must integrate the matrix element over momentum $\mathbf{K}_{f}$.
From (\ref{eq.app.2.5.5}) we obtain:
\begin{equation}
\begin{array}{ll}
  \langle \Psi_{f} |\, \hat{H}_{\gamma} |\, \Psi_{i} \rangle_{1} =
  -\,\displaystyle\frac{e}{m}\:
  \sqrt{\displaystyle\frac{2\pi\hbar}{w_{\rm ph}}}\;
    \displaystyle\sum\limits_{\alpha=1,2} \mathbf{e}^{(\alpha),*}\;
    \biggl\langle\: \Phi_{\rm \pi-nucl, f} (\mathbf{r})\: \biggl|\,
      Z_{\rm eff}(\mathbf{k}, \mathbf{r})\: e^{-i\,\mathbf{kr}}\: \mathbf{p}\;
    \biggr|\: \Phi_{\rm \pi-nucl, i} (\mathbf{r})\: \biggr\rangle, &
  \mathbf{K}_{i} = \mathbf{K}_{f} + \mathbf{k}.
\end{array}
\label{eq.app.2.5.8}
\end{equation}
%
%
%
The effective charge of the system in the first approximation of $\exp(i\mathbf{kr}) \to 1$ (i.~e. at $\mathbf{kr} \to 0$, called as \definition{dipole} concerned with the effective charge)
obtains form:
\begin{equation}
  Z_{\rm eff}^{\rm (dip)} (\mathbf{k}) =
  \displaystyle\frac{m_{A}\, z_{\pi} - m_{\pi}\, Z_{\rm A}(\mathbf{k})}{m_{A}+m_{\pi}}.
\label{eq.app.2.5.10}
\end{equation}
One can see independence of the effective charge on relative distance between $\pi^{\pm}$ and center-of-mass of the nucleus in such an approximation.

A simple determination of the matrix element can be obtained, it to neglect relative displacements of nucleons of the nucleus inside its space region in calculations of the form-factor (i.e. in approximation where the nucleus is considered as point-like and we use $e^{-i \mathbf{k} \rhobfsm_{Aj}} \to 1$ for each nucleon).
Here, the form-factor of the nucleus represents summarized electromagnetic charge of nucleons of the nucleus,
where dependence on characteristics of the emitted photon is lost:
%
\begin{equation}
  Z_{\rm A} (\mathbf{k}) \to
  \Bigl\langle \psi_{\rm nucl, f} (\rhobf_{A1} \ldots \rhobf_{AA-1})\: \Bigl|\;
    \displaystyle\sum\limits_{j=1}^{A}
      z_{Aj}\:
  \Bigr|\: \psi_{\rm nucl, i} (\rhobf_{A1} \ldots \rhobf_{AA-1})\, \Bigr\rangle =
  \displaystyle\sum\limits_{j=1}^{A}\, z_{Aj} = Z_{\rm A},
\label{eq.app.2.5.11}
\end{equation}
as the functions $\psi_{\rm nucl, s}$ are normalized.
At such approximations we obtain the matrix element (we add index \emph{(dip)}):
\begin{equation}
\begin{array}{ll}
  \langle \Psi_{f} |\, \hat{H}_{\gamma} | \Psi_{i} \rangle_{1}^{\rm (dip)} =
  -\displaystyle\frac{e}{m}\:
  \sqrt{\displaystyle\frac{2\pi\hbar}{w_{\rm ph}}}
  Z_{\rm eff}^{\rm (dip,0)}
    \displaystyle\sum\limits_{\alpha=1,2} \mathbf{e}^{(\alpha),*}
    \Bigl\langle\, \Phi_{\rm \pi-nucl, f} (\mathbf{r})\, \Bigl|\,
      e^{-i\,\mathbf{kr}}\: \mathbf{p}\;
    \Bigr|\: \Phi_{\rm \pi-nucl, i} (\mathbf{r})\: \Bigr\rangle, &
  Z_{\rm eff}^{\rm (dip,0)} =
  \displaystyle\frac{m_{A}\, z_{\pi} - m_{\pi}\, Z_{\rm A}}{m_{A}+m_{\pi}}.
\end{array}
\label{eq.app.2.5.12}
\end{equation}
Using as the functions $\psi_{\rm \pi-nucl, s} (\mathbf{r})$ the wave packets
(as in the formalism of Refs.~\cite{Maydanyuk.2003.PTP,Maydanyuk.2006.EPJA,Maydanyuk.2008.EPJA,Giardina.2008.MPLA,
Maydanyuk.2009.NPA,Maydanyuk.2009.TONPPJ,Maydanyuk.2009.JPS,Maydanyuk_Zhang_Zou.2016.PRC,Maydanyuk.2010.PRC,Maydanyuk.2011.JPCS}):
\begin{equation}
  \Phi_{\rm \pi-nucl, s} (\mathbf{r}, t) =
  \displaystyle\int\limits_{0}^{+\infty}
    g(k - k_{s})\:
    \psi_{\rm \pi-nucl, s} (\mathbf{r})\:
    e^{-iw(k)t}\; dk,
\label{eq.app.2.5.13}
\end{equation}
we rewrite the matrix element above (see Eq.~(6) in Ref.~\cite{Maydanyuk.2012.PRC} without spinor terms):
\begin{equation}
\begin{array}{lcl}
\vspace{1mm}
  \langle \Psi_{f} |\, \hat{H}_{\gamma} |\, \Psi_{i} \rangle_{1}^{\rm (dip)} =
  \displaystyle\frac{e}{m}\, \sqrt{\displaystyle\frac{2\pi\hbar}{w_{\rm ph}}} \cdot
  p_{\rm el} \cdot 2\pi\; \delta(w_{i} - w_{f} - w), \\

  p_{\rm el} =
  -\, Z_{\rm eff}^{\rm (dip,0)}\,
    \displaystyle\sum\limits_{\alpha=1,2} \mathbf{e}^{(\alpha),*}\;
    \Bigl\langle\: \psi_{\rm \pi-nucl, f} (\mathbf{r})\: \Bigl|\,
      e^{-i\,\mathbf{kr}}\: \mathbf{p}\;
    \Bigr|\: \psi_{\rm \pi-nucl, i} (\mathbf{r})\: \Bigr\rangle.
\end{array}
\label{eq.app.2.5.14}
\end{equation}

We calculate the matrix element in multipolar expansion of wave function of photons~\cite{Maydanyuk.alpha_decay.all}
%
\begin{equation}
\begin{array}{lcl}
  p_{\rm el} =
  i\, Z_{\rm eff}^{\rm (dip,0)}\,
  \sqrt{\displaystyle\frac{\pi}{2}}
  \displaystyle\sum\limits_{l_{\rm ph}=1}\,
    (-i)^{l_{\rm ph}}\, \sqrt{2l_{\rm ph}+1}\; \cdot
  \Bigl[ p_{l_{\rm ph}}^{M} - i\,p_{l_{\rm ph}}^{E} \Bigr], &
%
%
  p_{l_{\rm ph}}^{M} = \displaystyle\sum\limits_{\mu = \pm 1} h_{\mu}\, \mu\, p_{l_{\rm ph}\mu}^{M}, &
  p_{l_{\rm ph}}^{E} = \displaystyle\sum\limits_{\mu = \pm 1} h_{\mu}\, p_{l_{\rm ph}\mu}^{E},
\end{array}
\label{eq.app.app.1.3.4}
\end{equation}
where
\begin{equation}
\begin{array}{lcllcl}
  p_{l_{\rm ph}\mu}^{M} & = &
    \displaystyle\int
        \varphi^{*}_{f}(\mathbf{r}) \,
        \biggl( \displaystyle\frac{\partial}{\partial \mathbf{r}}\, \varphi_{i}(\mathbf{r}) \biggr) \,
        \mathbf{A}_{l_{\rm ph}\mu}^{*} (\mathbf{r}, M) \;
        \mathbf{dr}, &

  \hspace{7mm}
  p_{l_{\rm ph}\mu}^{E} & = &
    \displaystyle\int
        \varphi^{*}_{f}(\mathbf{r}) \,
        \biggl( \displaystyle\frac{\partial}{\partial \mathbf{r}}\, \varphi_{i}(\mathbf{r}) \biggr)\,
        \mathbf{A}_{l_{\rm ph}\mu}^{*} (\mathbf{r}, E)\;
        \mathbf{dr}.
\end{array}
\label{eq.app.app.1.3.2}
\end{equation}
Here, $\mathbf{A}_{l_{\rm ph}\mu}(\textbf{r}, M)$ and $\mathbf{A}_{l_{\rm ph}\mu}(\textbf{r}, E)$ are magnetic and electric multipolar terms
($j_{\rm ph}$ is quantum number characterizing eigenvalue of the full momentum operator, while $l_{\rm ph}= j_{\rm ph}-1, j_{\rm ph}, j_{\rm ph}+1$ is connected with orbital momentum operator, but it defines eigenvalues of photon parity).




\begin{thebibliography}{99}
\bibitem{Kluge.1991.RepPP}
  W.~Kluge,
\newblock
  \emph{Pion-nuclear scattering},
\newblock
  Rep. Prog. Phys. \textbf{54}, 1251--1332 (1991).

\bibitem{Hirata.1979.AP}
  M.~Hirata, J.~H.~Koch, E.~J.~Moniz, and F.~Lenz,
\newblock
  \emph{Isobar Hole Doorway States and pi O-16 Scattering},
\newblock
  Annals Phys. \textbf{120}, 205 (1979).

\bibitem{Oset.1982.PRep}
  E.~Oset, H.~Toki, and W.~Weise,
\newblock
 \emph{Pionic modes of excitation in nuclei},
\newblock
  Phys. Rept. \textbf{83}, 281 (1982).

\bibitem{Ericson_Weise.1988.book}
  T.~Ericson and W.~Weise,
\newblock
  ``Pions and Nuclei''
\newblock
  (Clarendon Press, Oxford, 1988).

\bibitem{Binon.1970.NPB}
  F.~Binon, P.~Duteil, J.~P.~Garron, 
  J.~Gorres, L.~Hugon, J.~P.~Peigneux, C.~Schmit, M.~Spighel, and J.~P.~Stroot,
\newblock
  \emph{Scattering of negative pions on Carbon},
\newblock
  Nucl. Phys. B \textbf{17}, 168 (1970).

\bibitem{Sobie.1984.PRC}
  R.~J.~Sobie, T.~E.~Drake, K.~L.~Erdman, 
  R.~R.~Johnson, H.~W.~Roser, R.~Tacik, E.~W.~Blackmore, D.~R.~Gill, S.~Martin,
  C.~A.~Wiedner, and T.~Masterson,
\newblock
  \emph{Elastic and inelastic scattering of 50~MeV pions from $^{12}{\rm C}$, $^{32}{\rm S}$, and $^{34}{\rm S}$},
\newblock
  Phys. Rev. C \textbf{30}, 1612--1621 (1984).

\bibitem{Oram.1981.NIM} 
  C.~J.~Oram, J.~B.~Warren, G.~M.~Marshall, and J.~Doornbos,
\newblock
  Nucl. Instrum. Methods \textbf{179}, 95 (1981).

\bibitem{Sobie.1984.NIM} 
  R.~J.~Sobie, T.~E.~Drake, B.~M.~Barnett, 
  K.~L.~Erdrnan, W.~Gyles, R.~R.~Johnson, H.~W.~Roser, R.~Tacik, E.~W.~Blackmore,
  D.~R.~Gill, S.~Martin, C.~A.~Wiedner, and T.~Masterson,
\newblock
  Nucl. Instrum. Methods \textbf{219}, 501 (1984).

\bibitem{Clough.1974.NPB}
  A.~S.~Clough, G.~K.~Turner, B.~W.~Allardyce, 
  C.~J.~Batty, D.~J.~Baugh, W.~J.~McDonald,
  R.~A.~J.~Riddle, L.~H.~Watson, M.~E.~Cage, G.~J.~Pyle, and G.~T.~A.~Squier,
\newblock
  \emph{Pion-nucleus total cross sections from 88 to 860 MeV},
\newblock
  Nucl. Phys. B \textbf{76}, 15--28 (1974).

\bibitem{Preedom.1981.PRC}
  B.~M.~Preedom, S.~H.~Dam, C.~W.~Darden~III, 
  R.~D.~Edge, D.~J.~Malbrough, T.~Marks,
  R.~L.~Burman, M.~Hamm, M.~A.~Moinester, R.~P.~Redwine, M.~A.~Yates,
  F.~E.~Bertrand, T.~P.~Cleary, E.~E.~Gross, N.~W.~Hill, C.~A.~Ludemann,
  M.~Blecher, K.~Gotow, D.~Jenkins, and F.~Milder,
\newblock
  \emph{Positive pion-nucleus elastic scattering at 30 and 50 MeV},
\newblock
  Phys. Rev. C \textbf{23}, 1134--1140 (1981).

\bibitem{Boyer.1981.PRC}
  K.~G.~Boyer, W.~B.~Cottingame, L.~E.~Smith, 
  S.~J.~Greene, C.~Fred~Moore,
  J.~S.~McCarthy, R.~C.~Minehart, J.~F.~Davis,
  G.~R.~Burleson, G.~Blanpied,
  C.~A.~Goulding, H.~A.~Thiessen, and C.~L.~Morris,
\newblock
  \emph{Pion inelastic scattering to the low-lying states in $^{42,44,48}{\rm Ca}$: Determination of the neutron all proton multipole matrix elements},
\newblock
  Phys. Rev. C \textbf{24}, 598--604 (1981).

\bibitem{Morris.1981.PRC}
  C.~L.~Morris,
  K.~G.~Boyer, C.~Fred~Moore, 
  C.~J.~Harvey, K.~J.~Kallianpur, I.~B.~Moore, P.~A.~Seidl,
  S.~J.~Seestrom-Morris, D.~B.~Holtkamp,
  S.~J.~Greene, and W.~B.~Cottingame,
\newblock
  \emph{Pion inelastic scattering to low-lying states in $^{12}{\rm C}$ and $^{40}{\rm Ca}$},
\newblock
  Phys. Rev. C \textbf{24}, 231--235 (1981).

\bibitem{Albanese.1980.NPA}
  J.~P.~Albanese, J.~Arvieux, J.~Bolger, 
  E.~Boschitz, C.~H.~Q.~Ingram, J.~Jansen, and J.~Zichy,
\newblock
  \emph{Elastic scattering of positive pions by $^{16}{\rm 0}$ between 80 and 340 MeV},
\newblock
  Nucl. Phys. A \textbf{350}, 301--331 (1980).

\bibitem{Preedom.1979.NPA}
  B.~M.~Preedom, R.~Corfu, J.-P.~Egger, 
  P.~Gretillat, C.~Lunke, J.~Piffaretti, E.~Schwarz, J.~Jansen, and C.~Perrin,
\newblock
  \emph{A systematic study of $\pi^{+}$ and $\pi^{-}$ inelastic scattering from $^{28}{\rm Si}$ in the region of the $\pi N (3,3)$ resonance},
\newblock
  Nucl. Phys. A \textbf{326}, 385--400 (1979).

\bibitem{Albanese.1979.NIM}
  J.~P.~Albanese, J.~Arvieux, E.~T.~Boschitz, 
  R.~Corfu, J.~P.~Egger,
  P.~Gretillat, C.~H.~Q.~Ingram, C.~Lunke, E.~Pedroni, C.~Perrin,
  J.~Piffaretti, L.~Pflug, E.~Schwarz, C.~Wiedner, and J.~Zichy,
\newblock
  \emph{The SIN high resolution pion channel and spectrometer},
\newblock
  Nucl. Instr. Methods. \textbf{158}, 363--370 (1979).

\bibitem{Boyer.1984.PRC}
  K.~G.~Boyer, W.~J.~Braithwaite, W.~B.~Cottingame, 
  S.~J.~Greene, L.~E.~Smith, C.~Fred Moore,
  C.~L.~Morris, H.~A.~Thiessen, G.~S.~Blanpied, G.~R.~Burleson,
  J.~F.~Davis, J.~S.~McCarthy, R.~C.~Minehart, and
  C.~A.~Goulding,
\newblock
  \emph{Pion elastic and inelastic scattering from $^{40,42,44,48}{\rm Ca}$ and $^{54}{\rm Fe}$},
\newblock
  Phys. Rev. C \textbf{29}, 182--194 (1984).

\bibitem{Khallaf.2000.PRC}
  S.~A.~E.~Khallaf and A.~A.~Ebrahim,
\newblock
  \emph{Analysis of $\pi^{\pm}$-nucleus elastic scattering using a local potential},
\newblock
  Phys. Rev. C \textbf{62}, 024603 (2000).

\bibitem{Khallaf.2002.PRC}
  S.~A.~E.~Khallaf and A.~A.~Ebrahim,
\newblock
  \emph{Elastic and inelastic scattering of pions from nuclei using an equivalent local potential},
\newblock
  Phys. Rev. C \textbf{65}, 064605 (2002).

\bibitem{Akhter.2001.JPG}    
  Md.~A.~E.~Akhter, Sadia~Afroze~Sultana, H.~M.~Sen~Gupta, and R.~J.~Peterson,
\newblock
  \textit{Local optical model studies of pion-nucleus scattering},
\newblock
  J. Phys. G \textbf{27}, 755--771 (2001).

\bibitem{Ebrahim.2011.BrJP}
  A.~A.~Ebrahim,
\newblock
  \emph{Cluster model analysis of pion elastic and inelastic scattering from $^{12}{\rm C}$},
\newblock
  Braz. J. Phys. \textbf{41}, 146--153 (2011).

\bibitem{Satchler.1992.NPA}
  G.~R.~Satchler,
\newblock
  Nucl. Phys. \textbf{A540}, 533 (1992).

\bibitem{Johnson.1996.AP}
  M.~B.~Johnson and G.~R.~Satchler,
\newblock
  \textit{Characteristics of local pion-nucleus potentials that are equivalent to Kisslinger-type potentials},
\newblock
  Ann. Phys. (N.~Y.) \textbf{248}, 134--169 (1996).

\bibitem{Pluiko.1987.PEPAN}
  V.~A.~Pluyko and V.~A.~Poyarkov,
\newblock
  Phys. El. Part. At. Nucl. \textbf{18} (2), 374--418 (1987).

\bibitem{Kamanin.1989.PEPAN}
  V.~V.~Kamanin, A.~Kugler, Yu.~E.~Penionzhkevich, et al.,
\newblock
  Phys. El. Part. At. Nucl. \textbf{20} (4), 743--829 (1989).

\bibitem{Amusia_Buimistrov.1987}
  M. Ya. Amusia, V. M. Buimistrov, B. A. Zon \textit{et al.},
\newblock
  \emph{Polaryzed bremsstrahlung emission of particles and atoms}
\newblock
  ({Nauka}, {Moskva}, 1987), 335~p.

\bibitem{Amusia.1990}
  M.~Ya.~Amusia,
\newblock
  \textit{Bremsstrahlung emission}
\newblock
  ({Energoatomizdat}, {Moskva}, 1990), 208~p.


\bibitem{Kopytin.1997.YF}
  I.~V.~Kopytin, M.~A.~Dolgopolov, T.~A.~Churakova, and A.~S.~Kornev,
\newblock
  Yad. Fiz. \textbf{60} (5), 869--879 (1997);
  Phys. At. Nucl. \textbf{60}, 776 (1997).

\bibitem{Gil.1998.PLB}
  A.~Gil and E.~Oset,
\newblock
  \emph{Coherent $\gamma$-production in $(p, p^{\prime})$ reactions in nuclei in the $\Delta$ resonance region},
\newblock
  Phys. Lett. \textbf{B 416}, 257--262 (1998).

\bibitem{Maydanyuk.2011.JPG}
  S.~P.~Maydanyuk,
\newblock
  \emph{Multipolar model of bremsstrahlung accompanying proton decay of nuclei},
\newblock
  Jour. Phys. \textbf{G38} (8), 085106 (2011) [16~p.],
\newblock
  1102.2067.

\bibitem{Maydanyuk.2012.PRC}
  S.~P.~Maydanyuk,
\newblock
  \emph{Model for bremsstrahlung emission accompanying interactions between protons and nuclei from low energies up to intermediate energies: Role of magnetic emission},
\newblock
  Phys. Rev. \textbf{C86}, 014618 (2012),
\newblock
  arXiv:1203.1498.

\bibitem{Maydanyuk_Zhang.2015.PRC}
  S.~P.~Maydanyuk and P.-M.~Zhang,
\newblock
  \emph{New approach to determine proton-nucleus interactions from experimental bremsstrahlung data},
\newblock
  Phys. Rev. C \textbf{91}, 024605 (2015).

\bibitem{Maydanyuk.2003.PTP}
  S.~P.~Maydanyuk and V.~S.~Olkhovsky,
\newblock
  \textit{Does sub-barrier bremsstrahlung in $\alpha$-decay of $^{210}\mbox{Po}$ exist?}
\newblock
  Prog. Theor. Phys. \textbf{109} (2), 203--211 (2003),
\newblock
  nucl-th/0404090.

\bibitem{Maydanyuk.2006.EPJA}
  S.~P.~Maydanyuk and V.~S.~Olkhovsky,
\newblock
  \textit{Angular analysis of bremsstrahlung in $\alpha$-decay},
\newblock
  Europ. Phys. Journ. \textbf{A28} (3), 283--294 (2006),
\newblock
  nucl-th/0408022.

\bibitem{Maydanyuk.2008.EPJA}
  G.~Giardina, G.~Fazio, G.~Mandaglio, 
  M.~Manganaro, C.~Sacc\'{a}, N.~V.~Eremin, A.~A.~Paskhalov, D.~A.~Smirnov, S.~P.~Maydanyuk, and V.~S.~Olkhovsky,
\newblock
  \textit{Bremsstrahlung emission accompanying  alpha-decay of $^{214}\mbox{\rm Po}$},
\newblock
  Europ. Phys. Journ. \textbf{A36} (1), 31--36 (2008);

\bibitem{Giardina.2008.MPLA}
  G.~Giardina, G.~Fazio, G.~Mandaglio, 
  M.~Manganaro, S.~P.~Maydanyuk, V.~S.~Olkhovsky, N.~V.~Eremin, A.~A.~Paskhalov, D.~A.~Smirnov, and C.~Sacc\'{a},
\newblock
  \textit{Bremsstrahlung emission during $\alpha$-decay of $^{226}\mbox{\rm Ra}$},
\newblock
  Mod. Phys. Lett. \textbf{A23} (31), 2651--2663 (2008),
\newblock
  arXiv:0804.2640;

\bibitem{Maydanyuk.2009.NPA}
  S.~P.~Maydanyuk, V.~S.~Olkhovsky, G.~Giardina, 
  G.~Fazio, G.~Mandaglio and M.~Manganaro,
\newblock
  Nucl.~Phys. \textbf{A823}, 3 (2009);

\bibitem{Maydanyuk.2009.TONPPJ}
  S.~P.~Maydanyuk,
\newblock
  \textit{Multipolar approach for description of bremsstrahlung during $\alpha$-decay and unified formula of the bremsstrahlung probability},
\newblock
  Open Nucl. Part. Phys. J. \textbf{2}, 17--33 (2009).

\bibitem{Maydanyuk.2009.JPS}
  S.~P.~Maydanyuk,
\newblock
  \textit{Multipolar approach for description of bremsstrahlung during $\alpha$-decay},
\newblock
  Jour. Phys. Study. \textbf{13} (3), 3201 (2009).

\bibitem{Maydanyuk_Zhang_Zou.2016.PRC}
  S.~P.~Maydanyuk, P.-M.~Zhang, and L.-P.~Zou,
\newblock
  \emph{New approach to determine proton-nucleus interactions from experimental bremsstrahlung data},
\newblock
  Phys. Rev. \textbf{C91}, 024605 (2016).

\bibitem{Maydanyuk.2010.PRC}
  S.~P.~Maydanyuk, V.~S.~Olkhovsky, G.~Mandaglio, 
  M.~Manganaro, G.~Fazio, and G.~Giardina,
\newblock
  \textit{Bremsstrahlung emission of high energy accompanying spontaneous of $^{252}{\rm Cf}$},
\newblock
  Phys. Rev. \textbf{C82}, 014602 (2010).

\bibitem{Maydanyuk.2011.JPCS}
  S.~P.~Maydanyuk, V.~S.~Olkhovsky, G.~Mandaglio, 
  M.~Manganaro, G.~Fazio, and G.~Giardina,
\newblock
  \emph{Bremsstrahlung emission of photons accompanying ternary fission of $^{252}{\rm Cf}$},
\newblock
  Journ. Phys.: Conf. Ser. \textbf{282}, 012016 (2011).


\bibitem{Ahiezer.1981}
  A.~I.~Ahiezer and V.~B.~Berestetskii,
\newblock
  \textit{Kvantovaya Elektrodinamika}
\newblock
  ({Nauka}, {Mockva}, 1981) p.~432 --- [in Russian].


\bibitem{Edington.1966.NP}
  J.~Edington and B.~Rose,
\newblock
  Nucl. Phys. \textbf{89}, 523 (1966).

\bibitem{Goethem.2002.PRL}
  M.~J.~van~Goethem, 
  L.~Aphecetche, J.~C.~S.~Bacelar, H.~Delagrange, J.~Diaz, D.~d'Enterria, M.~Hoefman,
  R.~Holzmann, H.~Huisman, N.~Kalantar-Nayestanaki, A.~Kugler, H.~L\"{o}hner, G.~Martinez,
  J.~G.~Messchendorp, R.~W.~Ostendorf, S.~Schadmand, R.~H.~Siemssen, R.~S.~Simon,
  Y.~Schutz, R.~Turrisi, M.~Volkerts, V.~Wagner, and H.~W.~Wilschut,
\newblock
  \textit{Suppresion of soft nuclear bremsstrahlung in proton-nucleus collisions},
\newblock
  Phys. Rev. Lett. \textbf{88} (12), 122302 (2002).

\bibitem{Becchetti.1969.PR}
  F.~D.~Becchetti, Jr. and G.~W.~Greenlees,
\newblock
  Phys. Rev. \textbf{182} (4), 1190--1209 (1969).


\bibitem{Clayton.1992.PRC}
  J.~Clayton, W.~Benenson, M.~Cronqvist, R.~Fox, D.~Krofcheck, R.~Pfaff,
  M.~F.~Mohar, C.~Bloch, D.~E.~Fields,
\newblock
  Phys. Rev. \textbf{C45}, 1815 (1992).

\bibitem{Clayton.1991.PhD}
  J.~E.~Clayton,
\newblock
  \emph{High energy gamma ray production in proton induced reactions
  at energies of 104, 145, and 195 MeV},
\newblock
  PhD thesis (Michigan State University, 1991).

\bibitem{Boie.2007.PRL}
   H.~Boie, H.~Scheit, U.~D.~Jentschura, F.~K\"{o}ck, M.~Lauer, A.~I.~Milstein, I.~S.~Terekhov, and D.~Schwalm,
\newblock
  \textit{Bremsstrahlung in $\alpha$ decay reexamined},
\newblock
  Phys. Rev. Lett. \textbf{99}, 022505 (2007).
\newblock
  arXiv:0706.2109.

\bibitem{Boie.2009.PhD}
  H.~Boie,
\newblock
  \textit{Bremsstrahlung emission probability in the $\alpha$ decay of $^{210}{\rm Po}$},
\newblock
  PhD thesis (Ruperto-Carola University of Heidelberg, Germany, 2009), 193~p.


%
\bibitem{Maydanyuk.2015.NPA}
  S.~P.~Maydanyuk, P.-M.~Zhang, and S.~V.~Belchikov,
\newblock
  \emph{Quantum design using a multiple internal reflections method in a study of fusion processes in the capture of alpha-particles by nuclei},
\newblock
  Nucl. Phys. A \textbf{940}, 89--118 (2015);
\newblock
  arXiv:1504.00567.

\bibitem{Maydanyuk_Zhang_Zou.2017.PRC}
  S.~P.~Maydanyuk, P.-M.~Zhang, and L.-P.~Zou,
\newblock
  \emph{New quasibound states of the compound nucleus in $\alpha$-particle capture by the nucleus},
\newblock
  Phys. Rev. \textbf{C96}, 014602 (2017);
\newblock
  arXiv:1711.07012.


\end{thebibliography}
\end{document}